\documentclass{article}
\usepackage{graphicx}
\usepackage{epstopdf, epsfig}

\title{First-order analysis of slip flow at the microscale and nanoscale}

\author{Duncan A. Lockerby (d.lockerby@warwick.ac.uk) \\ School of Engineering, University of Warwick
Coventry, CV4 7AL, UK}

\usepackage{geometry}
\usepackage{amssymb}
\usepackage{amsmath}
\usepackage{rsfso}
\usepackage{listings}
\usepackage{upgreek}
\usepackage{setspace}
\usepackage{graphicx}
\usepackage[textfont = sl, margin = 1.54mm, justification=centering]{caption}
\usepackage{subcaption}
\usepackage{multirow}
\usepackage[round]{natbib}

\usepackage{mathtools}
\usepackage{arydshln}
\usepackage{hhline}
\usepackage{amsfonts}
\usepackage{xcolor}
\newcommand\Kn{\mbox{\textit{K\hspace{-0.03cm}n}}} %
\newcommand{\bo}[1]{{\boldsymbol{#1}}}

\begin{document}

\maketitle

\abstract{A convenient approach to derive simple expressions for properties of Stokes flows with low levels of slip is presented. The method is based on a series expansion of a Stokes-flow solution (one satisfying a Navier slip boundary condition) with non-dimensional slip length as the small expansion parameter. Most notably, first-order predictions of surface moments of the traction force (e.g., drag and torque) can be obtained purely from no-slip solutions to the same problem. The analysis is directly applicable to microscale rarefied gas flows in the so-called `slip regime' and relevant to a range of liquid flows at the microscale and nanoscale. A number of application examples are considered, with expressions derived for: the drag and torque on translating and rotating Janus particles and spheroids (prolate and oblate); the efficiency of a micro journal bearing; the speed of a self-propelled particle (a `squirmer'); and the pressure drop required to drive flow through long, straight micro/nano channels. Where appropriate, accurate numerical calculations provide verification of the derived expressions. Certain general results are also obtained. For example, for low-slip Stokes flow: any surface distribution of positive slip length will reduce the drag on any translating particle; any perimetric distribution of positive slip length will reduce the pressure loss through a straight channel flow of arbitrary cross-section; unlike in no-slip flows, the rate of work done by a bounding solid surface on the fluid is not balanced by dissipation in the fluid volume --- there is additional dissipation at the fluid-solid interface.
}

\section{Introduction}
In microscale and nanoscale fluid mechanics, the relative tangential motion of a fluid and a solid at an interface, known as velocity slip, is a  familiar phenomenon. It is prominent, particularly, in microscale gas flows \citep{gad-el-hak_fluid_1999, karniadakis_micro_2002, arkilic_gaseous_1997} and in liquid flows at the microscale and nanoscale \citep{choi_apparent_2003, holt_fast_2006,  tropea_microfluidics_2007, falk_molecular_2010, qin_measurement_2011, nicholls_water_2012}. Typically, these flows are at very low Reynolds number, due to their scale, and this is assumed to be the case throughout this article. 

The standard approach to slip modelling is with a Maxwell or Navier slip boundary condition, for gases and liquids, respectively \citep{maxwell_stresses_1879,lockerby_velocity_2004, tropea_microfluidics_2007}. The two conditions are essentially equivalent, with both relating the velocity slip to the shear stress at the interface. In both cases, the degree of slip can be articulated using a slip length (described in detail later).

The focus of this paper is on \emph{low-slip flows}, i.e. flows for which the slip length is small relative to a characteristic scale of the flow geometry. In rarefied gas dynamics and for common surfaces, this is equivalent to the geometry being in the so-called `slip regime', for which the adoption of a Maxwell slip condition with the conventional continuum equations is the accepted model. For air at standard-atmospheric pressure, this corresponds to devices and particulate on the scale of microns. Liquid slip in micro and nano geometries is less understood and harder to predict, and so the scale that can be classified as \emph{low slip} is problem dependent. 

Super-hydrophobic surfaces generate slip by trapping pockets of gas within micro or nano structures at the liquid-solid interface \citep{rothstein_slip_2010}. This creates regions of very high slip (at the gas pockets) adjacent to regions of  no slip (at the structures). What is often done is to calculate an `effective' slip length for the heterogeneous surface \citep{lauga_effective_2003,belyaev_effective_2010}, which can be used with a Navier slip condition on a larger scale than the surface structures to predict their macroscopic impact. Cases where the effective slip length is small relatively to the macro geometry also fall into the category of {\emph{low slip}}; and when at very low Reynolds number they are within the scope of this work.

The motivation of this article is to present a method for deriving properties of low-slip Stokes flows in a general and convenient fashion. For example, we will show that in low-slip Stokes flow the drag on a single particle, of any shape, is well approximated by:
\begin{equation}\label{res1stOrderExpansion}
D \approx D_0-\tfrac{l}{ W \mu} \int_{S} {\tau}^2_{0}\, \, dS   \, ,
\end{equation}
where $D_0$ is the no-slip drag result,  $l$ is the slip length, $W$ is the speed of the particle, $\mu$ is the viscosity, and $\tau_0$ is the shear-stress magnitude over the particle surface ($S$) from the no-slip solution to the same problem. We will also show that the pressure difference ($\Delta p$) required to drive low-slip flow through straight channels, of any cross-sectional shape, can be calculated (note, very similarly) using: 
\begin{eqnarray}
\Delta p \approx \Delta p_0- \tfrac{ l \mathcal{L}}{Q \mu} \int_{\mathcal{P}} {\tau}_{0}^2\, \, d \mathcal{P} , 
\end{eqnarray}
where $\Delta p_0$ is the no-slip result, $\mathcal{L}$ is the channel length, $Q$ is the volumetric flow rate, and now the integral of $\tau_0^2$ is over the perimeter of the channel cross-section ($\mathcal{P}$).

Importantly, the approach allows simple analytical results to be derived for low-slip flows, from existing no-slip solutions, for cases where full slip solutions cannot be obtained or are extremely difficult to derive.

The paper is structured as follows. In \S\ref{sec:theory} we present the theoretical development and make brief observations on the implications of the main result. In \S\ref{sec:examples} we present a number of examples of its use, relevant to a range of microscale and nanoscale flow applications, including: predicting the mobility of particles with varying slip properties (e.g. Janus particles, \S\ref{sec:variableSurfaceSphere}) and with non-spherical geometry (e.g., prolate and oblate spheroids, \S\ref{sec:spheroids}); assessing the efficiency of a micro journal bearing (\S\ref{sec:journalBearing}); predicting the speed of a self-propelled `squirmer' with slip (\S\ref{sec:squirmer}); and evaluating pressure loss in flow through straight micro/nano channels (\S\ref{sec:channel}). In the context of low-slip flows, and with insight provided from \S\ref{sec:theory}, in \S\ref{sec:energy} we discuss the relationship between the rate of work done on the fluid by a bounding surface and the energy-dissipation rate within the fluid volume. The main result of this work also provide a means of \emph{numerically} estimating the impact of slip in Stokes flow, purely from the post-processing of no-slip solutions or numerical calculations. This is discussed in \S\ref{sec:discussion}, alongside other general comments.

\section{Theory}\label{sec:theory}
In this work, we restrict our attention to very low Reynolds number and steady-state flows, for which the governing equations are the steady Stokes equations:
\begin{eqnarray}\label{eq:contmom}
\bo{\nabla} \cdot \bo{u}=0 \, , \quad \, 
\bo{\nabla}\cdot \bo{\sigma}+\bo{f} =0\, ,
\end{eqnarray}
where $\bo{u}$ is the velocity, $\bo{f}$ is an applied body force, and $\bo{\sigma}$ is the stress tensor, composed of both pressure and viscous stresses:
\begin{eqnarray}\label{eq:consRel}
\bo{\sigma}=-p\bo{I}+ \mu \left({\bo{\nabla} \bo{u}}+\bo{\nabla} \bo{u}^T \right) \, ,
\end{eqnarray}
where $p$ is the pressure and $\mu$ is the dynamic viscosity.
Here we are concerned with solutions to the Stokes equations that satisfy a Navier slip condition at the boundary of the domain ($S$):
\begin{equation}\label{eq:navierslip}
    \bo{u}(\bo{r})=\bo{U}(\bo{r})+\frac{\ell(\bo{r})}{\mu}\,\, \bo{n}\cdot\bo{\sigma}\cdot(\boldsymbol{I}-\bo{n}\bo{n})\quad \mathrm{for} \quad \bo{r} \in S \, ,
\end{equation}
where $\bo{r}$ is the position vector, $\bo{u}$ is the velocity of the fluid at the fluid-solid interface, $\bo{U}$ is the velocity of the bounding surface itself (sometimes referred to as `the wall'),  $\bo{n}$ is a surface normal directed into the fluid, and $\ell(\bo{r})$ is a spatially varying slip length:
\begin{equation}\label{eq:varSlip}
\ell=l \, \psi(\bo{r})  \, ,
\end{equation}
where $l$ is the maximum slip length and $0\le \psi(\bo{r})\le 1$ is a non-dimensional function of position on the boundary. In most of the examples considered in this paper, surface properties are considered to be uniform, $\psi=1$, so that $ \ell\equiv l$. Note, Equation (\ref{eq:navierslip}) is general enough to represent velocity conditions at open boundaries (inlet/outlets) and zero-disturbance far-field conditions, by setting $\phi=0$, and specifying $\bo{U}$, accordingly.

The vector $\bo{n}\cdot\bo{\sigma}$
appearing in (\ref{eq:navierslip}) 
is the surface traction force, and its inner product with the tensor  $(\bo{I}-\bo{n}\bo{n})$ removes its surface-normal component, yielding a surface shear-stress vector, $\bo{\tau}$ (which slip velocity is proportional to). Note, when resolved in the surface-normal direction, equation (\ref{eq:navierslip}) corresponds to the impermeability condition.  

The key parameter of the current work is the non-dimensional maximum slip length\footnote{In cases where the distinction is unimportant, for brevity, we will sometimes refer to the `non-dimensional maximum slip length' as the `non-dimensional slip length', or just `slip length'}:
\begin{equation}
\xi=\frac{l}{L}  \, ,
\end{equation}
which expresses the degree of slip relative to a characteristic length scale of the flow in question, $L$.  In this paper, our attention is restricted to situations where the level of slip is low: $\xi \ll 1$. 

Equation (\ref{eq:navierslip}) is equivalent to Maxwell's slip boundary condition for isothermal rarefied flows \citep{maxwell_stresses_1879,lockerby_velocity_2004}; where $\ell=\lambda (2-\alpha)/\alpha$, $\lambda$ is the mean free path, and  $\alpha$ is the accommodation coefficient. For most practical surfaces $\alpha\approx 1$, which makes the non-dimensional slip length approximately equal to the Knudsen number:
\begin{equation}
\xi\approx \frac{\lambda}{L} = \Kn \, .
\end{equation}
Importantly, in adopting Maxwell's slip model it is already implied that  $\xi \approx\Kn \ll 1$. In other words, the assumption of $\xi\ll 1$ is consistent with the study of low-speed rarefied gas flows in the slip regime, for which equations (\ref{eq:consRel}) and (\ref{eq:navierslip}) are valid. For higher degrees of rarefaction, Knudsen layers and other gas-kinetic phenomena make their application unsuitable \citep{cercignani_mathematical_1969, sone_kinetic_2002,lockerby_modelling_2008, torrilhon_modeling_2016}.

\subsection{Series expansion for Stokes flow with slip}
We start by expanding the slip Stokes-flow solution in an infinite power series using the non-dimensional slip length as a small parameter, $\xi\ll 1$: 
\begin{equation}\label{uSeries}
    \bo{u}=\bo{u}_{0}+\xi\bo{u}_1 +\xi^2\bo{u}_2  \, + \, ... =\sum
_{k=0}^{\infty} \bo{u}_{k} \xi^k  \, ,
\end{equation}
\begin{eqnarray}\label{stressSeries}
    \bo{\sigma}= \bo{\sigma}_0 +\xi\bo{\sigma}_1 +\xi^2\bo{\sigma}_2 \, + \, ... 
= \sum
_{k=0}^{\infty} \bo{\sigma}_{k} \xi^{k} \, ,
\end{eqnarray} 
where $\bo{u}_k,\bo{\sigma}_k$ is a Stokes-flow solution associated with the $k$th order of the expansion. Substituting (\ref{uSeries}) and (\ref{stressSeries}) into the Navier slip condition (\ref{eq:navierslip}), and equating orders of $\xi$, yields the boundary conditions for the successive Stokes flow solutions in the expansion. The velocity field of the first solution ($\bo{u}_{0}$) satisfies:
\begin{equation}\label{bc0Series}
    \bo{u}_{0}(\bo{r})=\bo{U}(\bo{r})  \quad \mathrm{for} \quad \bo{r} \in S \, ;
\end{equation}
and therefore corresponds to the no-slip solution. Subsequent solutions in the series satisfy a velocity slip proportional to the shear-stress vector of the previous solution: 
\begin{eqnarray}\label{bcSeries}
    \bo{u}_{1}(\bo{r})&=&\tfrac{L}{\mu}\psi\,\bo{n}\cdot\bo{\sigma}_{0}\cdot(\boldsymbol{I}-\bo{n}\bo{n}) \nonumber\\
    \bo{u}_{2}(\bo{r})&=&\tfrac{L}{\mu}\psi\,\bo{n}\cdot\bo{\sigma}_{1}\cdot(\boldsymbol{I}-\bo{n}\bo{n})  \nonumber\\
    &...& \nonumber \\
    \bo{u}_{k}(\bo{r})&=&\tfrac{L}{\mu}\psi\,\bo{n}\cdot\bo{\sigma}_{k-1}\cdot(\boldsymbol{I}-\bo{n}\bo{n})
    \quad \mathrm{for} \quad \bo{r} \in S 
    \end{eqnarray}

\subsection{Moments of the traction force}
The main focus of this paper is on calculating surface moments of the traction force:
   \begin{equation}\label{eq:moment}
M = \int_{S} \bo{g}\cdot \bo{\sigma}\cdot \bo{n} \,\, dS\, ,
\end{equation}
where $\bo{g}(\bo{r})$ defines the moment in question. For example, if $S_p$ is the solid boundary of a particle, then
\begin{equation}
\bo{g}(\bo{r}) = \left\{\begin{array}{ll}
                  \bo{i}_x  &\mathrm{for}\quad \bo{r} \in S_p  \\
                  0               &\mathrm{for}\quad  \bo{r} \in S / S_p
                \end{array}  \right. 
    \end{equation}
produces a moment corresponding to the $x$-component of the total hydrodynamic force acting on the particle. If, as another example, the function is of the form:
\begin{equation}
\bo{g}(\bo{r}) = \left\{\begin{array}{ll}
                  \bo{r}\times\bo{i}_x\ &\mathrm{for}\quad \bo{r} \in S_p  \\
                  0               &\mathrm{for}\quad  \bo{r}  \in S / S_p
                \end{array}  \right. 
    \end{equation}
    the moment corresponds to the hydrodynamic torque on the particle about the $x$ axis.

Substitution of equation (\ref{stressSeries}) into (\ref{eq:moment}), yields a series representation of the traction-force moment:
    \begin{equation}\label{MSeries}
    M=M_{0}+\xi M_1 +\xi^2 M_2  \, + \, ... = \sum
_{k=0}^{\infty} M_{k} \xi^k  \, ,
\end{equation}
where 
\begin{equation}\label{eq:genMoment}
M_k= \int_{S} \bo{g} \cdot \bo{\sigma}_k \cdot  \bo{n} \, \, dS \, .
   \end{equation}

For low-slip flows, the first-order approximation is a good one, i.e.:
\begin{equation}
 M \approx M_{0}+\xi M_1 \, .
\end{equation}
where  $M_1$ is the \emph{first-order slip-correction} coefficient. The purpose of this paper is to present a simple and convenient way of obtaining $M_1$. 
\subsection{Finding the slip-correction coefficient, $M_1$ } \label{sec:findingM1}
Let $\bo{u}_0'$, $\bo{\sigma}_0'$ be a no-slip Stokes solution satisfying the boundary condition
\begin{equation}\label{uprime0}
    \bo{u}'_{0}(\bo{r})=\bo{g}(\bo{r})  \quad \mathrm{for} \quad \bo{r} \in S \, ,
\end{equation}
with a body force $\bo{f}'=0$. We will refer to this as the conjugate solution. 
Substituting (\ref{uprime0}) into (\ref{eq:genMoment}) gives:
\begin{equation}\label{firstOrderSCFa}
M_1= \int_{S} \bo{u}'_{0} \cdot \bo{\sigma}_1 \cdot  \bo{n} \, \, dS    \, .
   \end{equation}
From the reciprocal theorem, equation (\ref{firstOrderSCFa}) can be written:
   \begin{eqnarray}
M_1= \int_{S} \bo{u}_1 \cdot\bo{\sigma}_0' \cdot  \bo{n} \, \, dS  - \int_{V} \bo{f} \cdot \bo{u}'_{0} \, \, dV \, .
\end{eqnarray}
Now, upon substituting the boundary conditions for the first-order solution for velocity, $\bo{u}_1$,  from equation (\ref{bcSeries}),  we obtain
      \begin{eqnarray}
M_1= \tfrac{L}{ \mu} \int_{S}  \psi\,\bo{n}\cdot\bo{\sigma}_{0}\cdot(\boldsymbol{I}-\bo{n}\bo{n}) \cdot\bo{\sigma}_0' \cdot  \bo{n} \, \, dS - \int_{V} \bo{f} \cdot \bo{u}'_{0} \, \, dV\, .
\end{eqnarray}
Given that $(\boldsymbol{I}-\bo{n}\bo{n}) \cdot\bo{\sigma}_0'=(\boldsymbol{I}-\bo{n}\bo{n}) \cdot\bo{\sigma}_0' \cdot(\boldsymbol{I}-\bo{n}\bo{n})$, this becomes: 
    \begin{eqnarray}\label{firstOrderSCFmain1}
M_1=\tfrac{L}{\mu} \int_{S} \psi \, \bo{\tau}_{0}\cdot\bo{\tau}'_{0}\, \, dS- \int_{V} \bo{f} \cdot \bo{u}'_{0} \, \, dV\, .
\end{eqnarray}
where  $\bo{\tau} = \bo{n}\cdot\bo{\sigma}\cdot(\boldsymbol{I}-\bo{n}\bo{n})$ is the tangential shear-stress vector.
Importantly, the right-hand side of equation (\ref{firstOrderSCFmain1}) is purely in terms of no-slip solutions. 

In all of the examples we will consider in this paper, there is no applied body force, and so we will omit the second term in (\ref{firstOrderSCFmain1}), and work with the simpler and normalised expression for the first-order slip-correction coefficient:
\begin{eqnarray}\label{firstOrderSCFmain}
\hat{M}_1=\tfrac{L}{M_0 \mu} \int_{S} \psi \, \bo{\tau}_{0}\cdot\bo{\tau}'_{0}\, \, dS ,
\end{eqnarray}
where $\hat{M}_1=M_1/M_0$. From hereon, hats denote a dimensionless value normalised with its corresponding no-slip quantity.
\subsection{The resistive moment, $\mathcal{R}$}\label{sec:resmom}
Equation (\ref{firstOrderSCFmain}) represents the main result of this article, but we now introduce an important special case, which simplifies (\ref{firstOrderSCFmain}) further. For most applications that we consider, the function $\bo{g}$ that defines the moment of the traction force has the same dependence on position as the velocity of the wall, but the opposite sign:
\begin{equation}\label{resMom1}
 \bo{U}(\bo{r})=-k \, \bo{g}(\bo{r}) \, 
\end{equation} 
where $k$ is a constant of proportionality. 
For example, we might want to calculate the $x$-component of drag on a particle ($\bo{g}=\bo{i}_x$) due to its translation in the opposite direction ($\bo{U}=-U \bo{i}_x$). In this case, $k=U$, where $U$ is the particle speed. Alternatively, we might want to calculate the torque around the $x$-axis of a particle ($\bo{g}=\bo{r}\times\bo{i}_x$) that rotates about the $x$-axis in the opposite sense ($\bo{U}=-\omega \, \bo{r}\times\bo{i}_x$). Here $k=\omega$, where $\omega$ is the magnitude of the angular velocity. 

We refer to this particular moment as the \emph{resistive moment}, using the symbol $\mathcal{R}$ to distinguish it from the general case. The resistive moment can be expanded, to first order in slip length, as previously:
\begin{equation} \label{eq:Rseries}
\frac{\mathcal{R}}{\mathcal{R}_0}=1+\hat{\mathcal{R}}_1 \xi + \mathcal{O}(\xi^2)
\end{equation}
where $\hat{\mathcal{R}}_1=\mathcal{R}_1/\mathcal{R}_0$, and the corresponding no-slip moment is given by:
\begin{equation}\label{eq:rmR0}
    \mathcal{R}_0=-\frac{1}{k} \int_{S} \bo{u}_{0} \cdot \bo{\sigma}_0 \cdot  \bo{n} \, \, dS    \, ,
\end{equation}
which in the absence of external forces is guaranteed to be positive.
To find $\mathcal{R}_1$, we substitute (\ref{bc0Series}) and (\ref{uprime0}) into (\ref{resMom1}) to give: 
\begin{equation}
 \bo{u}_0=-k \, \bo{u}'_0 \, 
\end{equation}
from which it follows that
\begin{equation}\label{tauPrime}
 \bo{\tau}_0=-k \, \bo{\tau}'_0 \, .
\end{equation}

Finally, substitution of (\ref{tauPrime}) into the general expression (\ref{firstOrderSCFmain}) gives the first-order slip-correction coefficient for the resistive moment:
  \begin{equation}\label{resCorrFac}
\hat{{\mathcal{R}}}_{1}= -\tfrac{L}{\mathcal{R}_0 k  \mu} \int_{S} \psi \, {\tau}^2_{0}\, \, dS \, ,
\end{equation}
where $\tau_0$ is the magnitude of the tangential shear stress. Note, for the resistive moment there is no additional conjugate no-slip solution.

It is significant that this is necessarily negative for any distribution of positive slip length.  Since $\mathcal{R}_0>0$, this tells us that a small amount of slip on any geometry will reduce the resistive moment of the traction force. In short, and for example, low levels of slip will reduce the drag on any translating geometry in Stokes flow (in the absence of applied body forces). Similarly, any distribution of slip length (provided it is small) will reduce the retarding torque on a rotating particle.

\section{Application examples and analytical solutions}\label{sec:examples}
The remainder of this paper is dedicated to applying the expressions obtained above to different flow problems, in order to derive analytical expressions for first-order slip corrections. Mostly, these are cases for which full analytical slip solutions either do not exist or are extremely involved to evaluate. 
\subsection{A sphere}\label{sec:simpleSphere}
\begin{figure}
    \centering
    \includegraphics[width=0.35\linewidth]{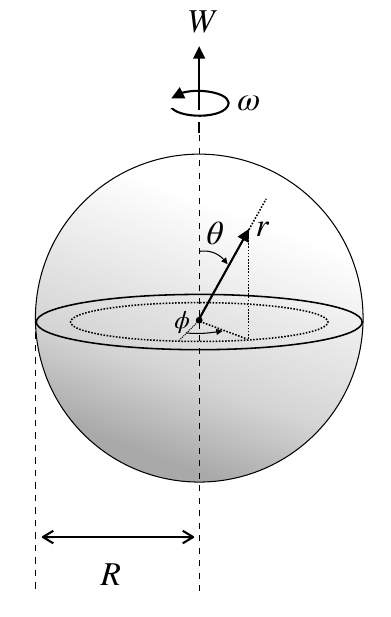}
    \caption{A translating and rotating sphere in spherical coordinates.}
    \label{fig:sphere}
\end{figure}

We start, though, by verifying the expressions derived in \S\ref{sec:theory} for a problem that has a simple and well-established analytical treatment for slip flow; namely, translational and rotational motion of a single sphere in free space; due to \cite{basset_treatise_1888}. Basset's solutions for drag and retarding torque, derived for a uniform slip length ($\psi=1$), are
\begin{equation}\label{eq:basset}
    D=D_0\left( \frac{1+2 \xi}{1+3\xi}\right)  \quad \mathrm{and}   \quad T=T_0 \left(\frac{1}{1+3\xi}\right) \quad 
\end{equation}
where $\xi=l/R$ is the non-dimensional slip length, $R$ is the sphere radius, $D_0=6\pi \mu  W R $ is the no-slip result for drag on a translating sphere with velocity $W$, and $T_0=8\pi \mu  \omega R^3$ is the no-slip result for retarding torque on a rotating sphere with angular velocity $\omega$; see Figure \ref{fig:sphere}. Expanding Basset's expressions in a Taylor series, gives the drag and torque to first order in slip length:
\begin{equation}\label{eq:spherefirstOrderCoeffs}
  \hat{D}=1-\xi + \mathcal{O}(\xi^2) \quad \mathrm{and}   \quad \hat{T}=1-3 \xi + \mathcal{O}(\xi^2) \, ,
\end{equation}
where $\hat{D}=D/D_0$ and $\hat{T}=T/T_0$. The first-order slip-correction coefficients are, therefore, $\hat{D_1}=-1$ and $\hat{T_1}=-3$ . 

A side note: the first-order slip-correction coefficient for drag ($\hat{D_1}=-1$) was first obtained from kinetic theory for dilute gas flows  by \cite{epstein_resistance_1924}, who also demonstrated that Basset's full-slip solution (\ref{eq:basset}) was only valid to this order; see \cite{happel_low_1983} for more discussion.

We now demonstrate how to obtain the first-order coefficients in equation (\ref{eq:spherefirstOrderCoeffs}) directly from the corresponding no-slip solutions. For the case of a translating sphere, and in spherical polar coordinates, the distribution of wall shear-stress magnitude in the no-slip solution is:
\begin{equation}\label{eq:noslipTranslatingSphereTau}
    \tau_0=\frac{3 \mu  W \sin \theta }{2 R} \, ,
\end{equation}
where $\theta$ is the polar angle (see Figure \ref{fig:sphere}). In this case, drag force is the resistive moment, and so the first-order slip-correction coefficient can be obtained directly from Equation (\ref{resCorrFac}) (with $\mathcal{R}_1= D_1$, $L=R$,  $\psi=1$, and $k = W$):
  \begin{equation}\label{eq:sprm1}
\hat{D}_1= -\tfrac{R}{D_0 W \mu} \int_{S} \, {\tau}^2_{0}\, \, dS=- \tfrac{9\pi \mu W R}{2  D_0} \int^\pi_0 \, \sin^3\theta\, \, d\theta=-1 \, ,
\end{equation}
which agrees with Basset and Epstein's solutions.

For the rotating sphere, the surface shear-stress magnitude with no slip is 
\begin{equation}\label{eq:noslipRotatingSphereTau}
    \tau_0=3 \mu  \omega  \sin \theta \, .
\end{equation}
For this case, retarding torque is the resistive moment, and so, again, the first-order slip-correction coefficient is obtained directly from Equation (\ref{resCorrFac}) (but with  $\mathcal{R}_1= T_1$,  $L=R$, $\psi=1$, and $k=\omega$):
 \begin{equation}\label{eq:sprm2}
\hat{T}_1= -\tfrac{R}{T_0 \omega \mu} \int_{S} \, {\tau}^2_{0}\, \, dS=- \tfrac{18\pi \mu \omega  R^3}{ T_0} \int^\pi_0 \, \sin^3\theta\, \, d\theta=-3\, ,
\end{equation}
which, again, agrees with the result of Basset.

\subsection{A sphere with varying slip length}\label{sec:variableSurfaceSphere}
We now choose an example for which full-slip solutions, like those due to Basset, do not exist. Consider a rigid sphere, as in Figure \ref{fig:sphere} and \S\ref{sec:simpleSphere}, but with a non-constant slip length: 
\begin{equation}
 \ell=l \,  \psi(\theta,\phi)  \, ,
\end{equation}
where $\psi$ is some arbitrary surface function and $l$ is the maximum slip length over the surface.
The first-order slip-correction coefficient for drag force due to translation comes directly from Equation (\ref{resCorrFac}) for the resistive moment (with $\mathcal{R}_1= D_1$, $L=R$, and $k = W$):
 \begin{equation}\label{eq:resMomVarSphere}
\hat{D}_1= - \tfrac{R}{D_0 W \mu} \int_{S} \psi \, {\tau}^2_{0}\, \, dS  \, ,
\end{equation}
where, as a reminder, $\tau_0$ is the surface shear-stress magnitude of a sphere in translation with no slip (Equation \ref{eq:noslipTranslatingSphereTau}). We can express $\tau_0^2$ in spherical harmonics:

\begin{equation}
\tau_0^2=\frac{3\sqrt{\pi} \mu^2 W^2 }{R^2}\left( Y^0_{0}-\frac{1}{\sqrt{5}}Y^0_{2} \right) \, ,   
\end{equation}
where $Y^0_0=\frac{1}{2\sqrt{\pi}}$  and $Y^0_2=
\frac{1}{4} \sqrt{\frac{5}{\pi }} \left(3 \cos ^2(\theta )-1\right)
$. On substitution into (\ref{eq:resMomVarSphere}), and then (\ref{eq:Rseries}), we obtain an expression for the drag on a variable-slip-length sphere:
  \begin{equation}
\hat{D}=1 -\bar{\psi}\xi +\frac{2\sqrt{5 \pi} }{ 5}   \psi^0_{2} \xi +\mathcal{O}(\xi^2)
\end{equation}
where  $\bar{\psi}\xi$ is the average slip length, $\bar{\psi}= \int_S \psi \, dS/(4\pi R^2)$  and $\psi^0_2 =\int_S \psi Y_2^0 \, dS/(4\pi R^2)$.

This is a surprising result: for the same average slip length, only variations in slip length that are of the form of the second zonal spherical harmonic ($ Y_2^0$) influence the drag in low-slip flow;  we can say this because of the orthogonality of spherical harmonics.  

\begin{figure}
    \centering
    \includegraphics[width=0.67\linewidth]{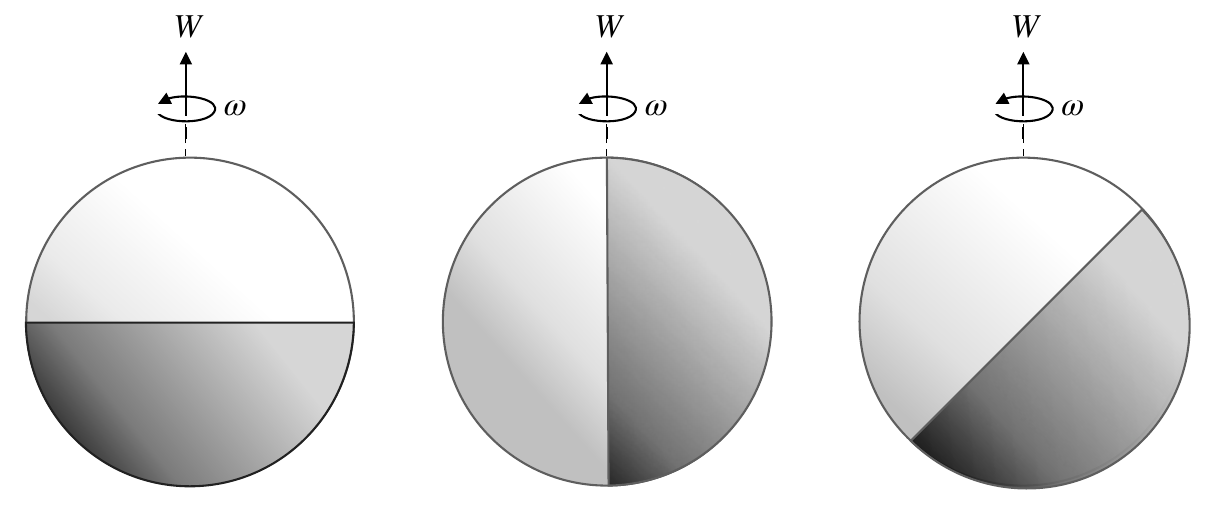}
    \caption{A type of Janus particle: a sphere with constant slip length on one hemisphere (dark grey) and no slip on the other (light grey); various orientations relative to the direction of translation and rotation.}
    \label{fig:janusSpheres}
\end{figure}

Let us take, for example, a type of Janus particle: a sphere having a constant slip length on one hemisphere and no slip on the other; see Figure \ref{fig:janusSpheres}. Irrespective of the orientation, the first-order slip-correction coefficient is half that of a sphere with a uniform slip length, since in all orientations $\bar{\psi}=\frac{1}{2}$ and $\psi^0_2=0$. In short, in low-slip flow, this type of particle has equal drag in all directions. 

\begin{figure}
    \centering
    \includegraphics[width=0.69\linewidth]{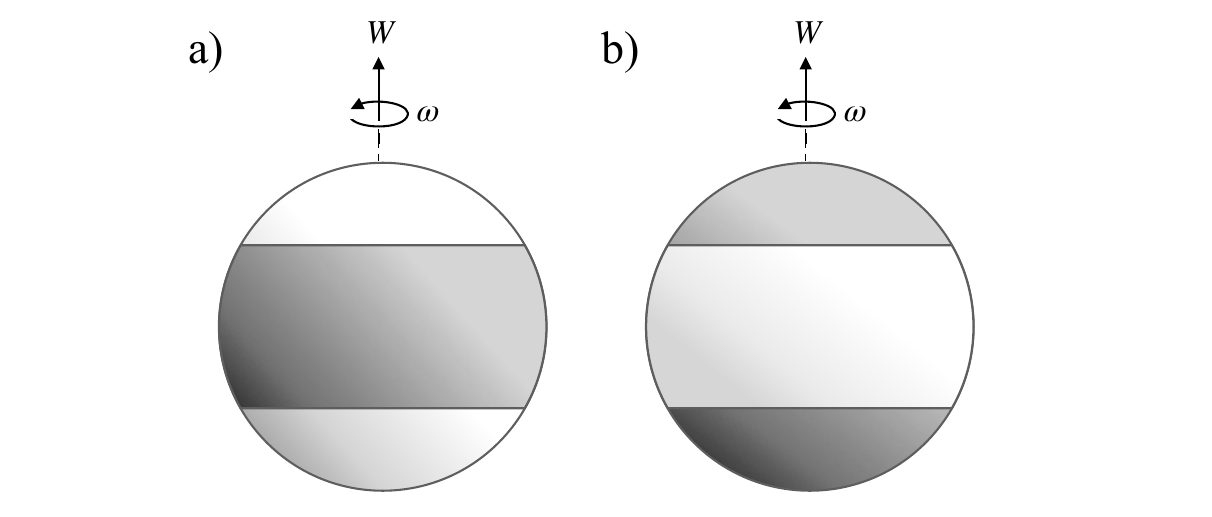}
    \caption{A sphere with: a) a constant slip length on a central band (dark grey) and no slip on the polar caps (light grey); and b) vice versa.}
    \label{fig:Janus2}
\end{figure}

A more interesting example is shown in Figure \ref{fig:Janus2}, where the slip (or no-slip) region is a central band, again, covering half the surface area. 
  For the case of a central band of constant slip length, Figure \ref{fig:Janus2}a,  $\bar{\psi}=\frac{1}{2}$ and  $\psi^0_2=-3 \sqrt{({5}/{\pi }) }/32$. For the case of polar regions of constant slip length, Figure \ref{fig:Janus2}b,  $\psi^0_0=\frac{1}{2}$  and $\psi^0_2=3 \sqrt{({5}/{\pi }) }/32$. 
For translation along the polar axis, the drag to first order in slip length is therefore given by: 
  \begin{equation}\label{eq:fvs1}
\hat{F}=1 -\frac{1}{2}\xi + \frac{3\upsilon}{16}\xi+  \mathcal{O}(\xi^2) \, .
\end{equation}
where $\upsilon=-1$ and $+1$ for the central-band and polar-regions case, respectively. In both cases, translation in a perpendicular direction to that shown in Figure \ref{fig:Janus2} results in a drag force given by (\ref{eq:fvs1}) with $\upsilon=0$ ($\bar{\psi}=\frac{1}{2}$ and $\psi^0_2=0$). As such, for the central-band case, the translation direction with minimum drag is in the direction shown in Figure \ref{fig:Janus2}a (i.e. perpendicular to the band). However, for the sphere with polar regions of slip, the translation direction with lowest drag is perpendicular to that indicated in Figure \ref{fig:Janus2}b.
   
An identical analysis can be repeated for the retarding torque due to rotation about the polar axis (see Figure \ref{fig:Janus2}). In the general case:
 \begin{equation}\label{eq:restorvarsph}
\hat{T}=1 -3 \bar{\psi}\,\xi+\frac{6\sqrt{5 \pi} }{ 5}   \psi^0_{2}\, \xi +\mathcal{O}(\xi^2)\, . 
\end{equation}
As with the drag-force case, the retarding torque in low-slip flow is only affected by the average slip length and slip-length variations in the form of the second zonal spherical harmonic. For the examples illustrated in Figure \ref{fig:Janus2}, 
\begin{equation}\label{eq:tvs1}
\hat{T}=1 -\frac{3}{2}\xi+\frac{9 \upsilon}{16}\xi+  \mathcal{O}(\xi^2) \, .
\end{equation}

To verify these analytical results, we perform numerical simulations of Stokes flow with a Navier slip condition, using the Method of Fundamental Solutions. Appendix \ref{sec:MFS} gives full details of the numerical methodology and the parameters used.

Tables \ref{table:vs1} and \ref{table:vs2} compare the numerical simulations to the analytical results for the cases illustrated in Figure \ref{fig:Janus2}. Other than to verify the derivations for low-slip conditions, the purpose of the comparison is to illustrate the extent to which predictions from the low-slip assumption diverge from the full-slip numerical simulation with increasing $\xi$. The analytical predictions are within 0.3\% of the numerical solutions for $\xi=10^{-4}$, and within 10\% for $\xi=10^{-2}$.

 \begin{table}
  \begin{spacing}{1.15}
\centering
\begin{tabular}{c c c c | c}
\hline \hline
& \multicolumn{3}{c|}{Numerical}  &   Eqn (\ref{eq:fvs1})\\
  &  $\xi=10^{-2}$ & $\xi=10^{-3}$ & $\xi=10^{-4}$& $\xi\to0$   \\
\hline
  Band&  $ -0.6495$& $ -0.6840$& $-0.6878$& $-0.6875$ \\
 Polar& $-0.2846$&$-0.3087$&$-0.3115$& $-0.3125$\\
\hline
\end{tabular}
\end{spacing}
\caption{Numerical and theoretical predictions for the slip correction: $(\hat{F}-1)/\xi $}\label{table:vs1}
\end{table}
 \begin{table}
  \begin{spacing}{1.15}
\centering
\begin{tabular}{c c c c | c}
\hline \hline
& \multicolumn{3}{c|}{Numerical}  &   Eqn (\ref{eq:tvs1})\\
  &  $\xi=10^{-2}$ & $\xi=10^{-3}$ & $\xi=10^{-4}$& $\xi\to0$   \\
\hline
Band&  $ -1.9723$& $-2.0547$& $-2.0636$& $-2.0625$ \\   
 Polar& $-0.8767$&$-0.9290$&$-0.9348$& $-0.9375$\\
 \hline
\end{tabular} 
\end{spacing}
\caption{Numerical and theoretical predictions for the slip correction: $(\hat{T}-1)/\xi $}\label{table:vs2}
\end{table}

\subsection{Prolate and Oblate Spheroids}\label{sec:spheroids}
Stokes flow along the axis-of-revolution of a no-slip spheroid has been solved, analytically, with a number of approaches \citep{oberbeck_ueber_1876,payne_stokes_1960, happel_low_1983}.
The problem of slip flow is substantially more complex. \cite{keh_slow_2008} dedicated a full article to the derivation, involving an infinite-series form of semi-separation of variables; truncating the series after two terms still requires a page of algebra to define the analytical coefficients.

Here, by contrast, we aim for a short and simple closed-form expression for the drag on a spheroid in low-slip conditions. This is a special case of that presented by \cite{keh_slow_2008}, but which was not derived/presented there.

\begin{figure}
    \centering
    \includegraphics[width=0.75\linewidth]{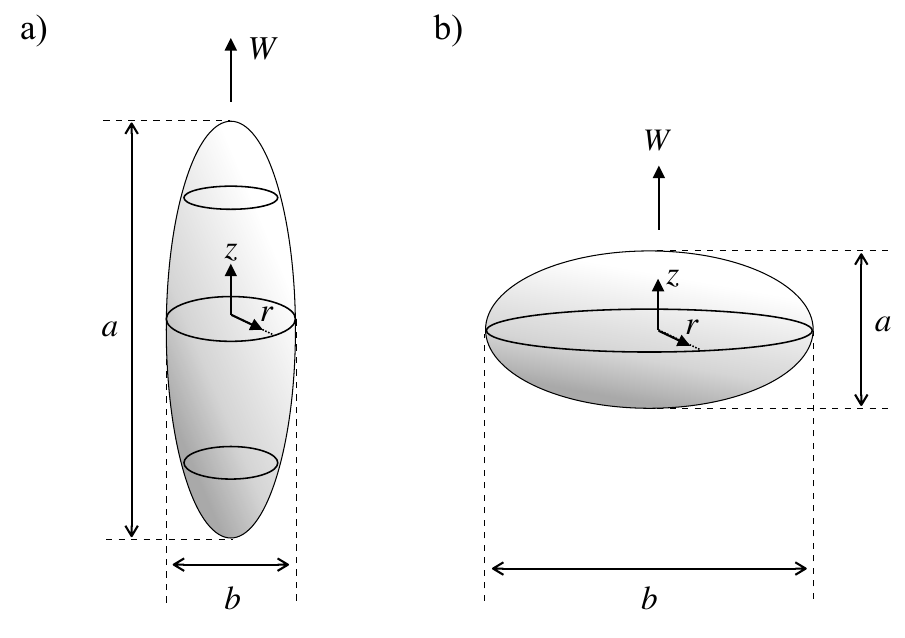}
    \caption{A spheroid in axial translation: a) prolate and b) oblate.  }
    \label{fig:spheroids}
\end{figure}

The implicit equation for the surface of a spheroid in cylindrical polar coordinates is:
 \begin{equation}
 \frac{r^2}{b^2}+\frac{z^2}{a^2} =1 \, ,
 \end{equation}
 where $b$ is the spheroid's equatorial radius and $a$ is the distance from centre to either pole; see Figure\,\ref{fig:spheroids}.
\subsubsection{Prolate spheroids, $a>b$}
The no-slip drag on a prolate spheroid in axial translation is given by \citep{payne_stokes_1960,sherman_viscous_1990}:
\begin{equation}
   D_0=\frac{16 \pi  \mu  \mathcal{L} W}{\left(s^2+1\right) \log \left(\frac{s+1}{s-1}\right)-2 s} \, ,
\end{equation}
where $W$ is the translational velocity of the spheroid in the direction of $z$,  $s={E}/{\sqrt{E^2-1}}$,  $E=a/b$,  and $\mathcal{L}=\sqrt{a^2-b^2}$ is the focal length.

The shear-stress magnitude from the no-slip solution is given by:
\begin{equation}\label{eq:pstau}
\tau_0=-\frac{4 E \mu  W \sin (\eta )}{\mathcal{L} \left(\cos (2 \eta )-2 s^2+1\right) \left(\left(s^2+1\right) \log \left(\coth \left(\frac{1}{2} \cosh ^{-1}(s)\right)\right)-s\right)}
\end{equation}
where $\eta$ is a coordinate on the the spheroid surface related to cylindrical polar coordinates through $z=\mathcal{L}\,s \cos(\eta) $ and $r=\mathcal{L} s \sin (\eta )/E$. 

The slip-flow drag on the prolate spheroid is approximated to first order by:
\begin{equation}\label{eq:spheroidEx}
D/D_0=1+\hat{D_1} \xi +\mathcal{O}(\xi^2)
\end{equation}
where $\xi=l/b$. For translation, darg is the resistive moment, and so the first-order slip-correction coefficient is obtained by evaluating equation (\ref{resCorrFac}) (with $\mathcal{R}= D$, $L=b$,  $\psi=1$, and $k = W$):
\begin{equation}\label{eq:prosc}
\hat{D}_1= -\tfrac{b}{D_0 U \mu} \int_{S} \, {\tau}^2_{0}\, \, dS=-\frac{2\left(1-  s E \tan ^{-1}\left(E/s\right)\right)}{ E  \left(1-\left(s+1/s\right) \coth ^{-1}(s)\right)}\, .
\end{equation}
Given the complexity of the full-slip derivation and solution due to \cite{keh_slow_2008}, equation (\ref{eq:prosc}) is remarkably simple.

In Table \ref{table:pro}, numerical calculations using the Method of Fundamental Solutions (see Appendix \ref{sec:MFS}) are compared to (\ref{eq:prosc}) for a range of spheroid aspect ratios ($E$); as expected, the full-slip numerical results converge to the analytical solution as the slip length is reduced. For $\xi=10^{-2}$ the analytical result is within $3\%$ of the numerical solutions; for $\xi=10^{-5}$ the analytical result is within 0.02\% of the numerical solutions.


\begin{table}
 \begin{spacing}{1.15}
\centering
\begin{tabular}{c |c c c  c | c }
\hline \hline 
& \multicolumn{4}{|c|}{Numerical $(D/D_0-1)/\xi$}  &   Eq.\,(\ref{eq:prosc})\\
$E$   &  $\xi=10^{-2}$ & $\xi=10^{-3}$ & $\xi=10^{-4}$ & $\xi=10^{-5}$& $\xi\to0$   \\
\hline
1.1 & $-$0.9546 & $-$0.9780 & $-$0.9805  & $-$0.9807 &  $-$0.9807 \\
1.5 &  $-$0.8965  & $-$0.9137  & $-$0.9154 &$-$0.9156 &$-$0.9156 \\
2 &$-$0.8388 &$-$0.8523  & $-$0.8537  & $-$0.8539&$-$0.8538\\
3 &$-$0.7566 & $-$0.7669  & $-$0.7680 & $-$0.7681 &$-$0.7681 \\
4 &$-$0.7004 & $-$0.7093  & $-$0.7102 &$-$0.7103 &$-$0.7104\\
\hline 
\end{tabular}
\end{spacing}
\caption{Numerical and analytical results for the prolate spheroid}\label{table:pro}
\end{table}
\subsubsection{Oblate spheroids, $a<b$}
The no-slip solution for drag on the oblate spheroid is \citep{payne_stokes_1960, sherman_viscous_1990}:
\begin{equation}
D_0=\frac{8 \pi  \mu  \mathcal{L} W}{t-(t^2-1)\cot ^{-1}(t)}\, ,
\end{equation}
where $t=\sinh(\cosh^{-1}(s))$ 
and the corresponding shear-stress magnitude distribution is:
\begin{equation}
\tau_0=\frac{4 \mu  W \sin (\eta )}{\mathcal{L} E \left(t-\left(t^2-1\right) \cot ^{-1}(t)\right) \left(\cos (2 \eta )+\cosh \left(2 \sinh ^{-1}(t)\right)\right)}\, ,
\end{equation}
where, now, $E=b/a$, $\mathcal{L}=\sqrt{b^2-a^2}$ and $\eta$ is a coordinate on the surface of the spheroid related to cylindrical polar coordinates through $z=\mathcal{L}\,t \cos(\eta) $ and $r=\mathcal{L} s \sin (\eta )$. 

The drag on the oblate spheroid in slip flow can be expanded, as in equation (\ref{eq:spheroidEx}), but with $\xi=l/a$, and the first-order slip-correction coefficient obtained from (\ref{resCorrFac}) (with $\mathcal{R}_i= D_i$, $L=a$,  $\psi=1$, and $k = W$):
\begin{equation}\label{eq:oblsc}
\hat{D}_1=-\tfrac{a}{D_0 U \mu} \int_{S} \, {\tau}^2_{0}\, \, dS=-\frac{2   \left( E - t \coth ^{-1}(s)\right)}{  E \left(1-(t-1/t)\cot ^{-1}(t)\right)}\, .
\end{equation}

As for the prolate case, the simplicity of the derivation and the final result is noteworthy. In Table \ref{table:obl}, numerical calculations (see Appendix \ref{sec:MFS}) verify equation (\ref{eq:oblsc}) and also provide an indication of the loss in accuracy of the low-slip assumption as the slip length increases.

\begin{table}
 \begin{spacing}{1.15}
\centering
\begin{tabular}{c | c c c  c | c }
\hline \hline
& \multicolumn{4}{|c|}{Numerical $(D/D_0-1)/\xi$}  &   Eq.\,(\ref{eq:oblsc})
\\
$E$   &  $\xi=10^{-2}$ & $\xi=10^{-3}$ & $\xi=10^{-4}$ & $\xi=10^{-5}$& $\xi\to0$   \\
\hline
1.1 & $-$0.8991  & $-$0.9234  & $-$0.9259  & $-$0.9262 & $-$0.9262 \\
1.5 & $-$0.6951  & $-$0.7150  & $-$0.7171  & $-$0.7173 &   $-$0.7173 \\
 2  & $-$0.5417  & $-$0.5591  & $-$0.5609  & $-$0.5611 &   $-$0.5611 \\
 3  & $-$0.3745  & $-$0.3898  & $-$0.3914  & $-$0.3916 &   $-$0.3916 \\
 4  & $-$0.2848  & $-$0.2991  & $-$0.3006  & $-$0.3008 &   $-$0.3007 \\
\hline
\end{tabular}
\end{spacing}
\caption{Numerical and analytical results for the oblate spheroid}
\label{table:obl}
\end{table}

\subsection{Journal Bearing}\label{sec:journalBearing}

\begin{figure}
    \centering
    \includegraphics[width=0.6\linewidth]{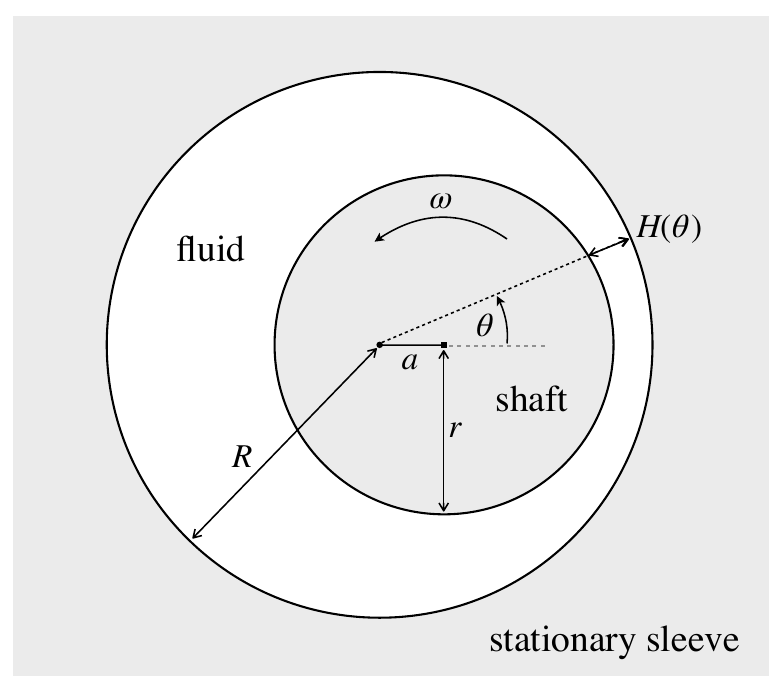}
    \caption{Schematic of a journal bearing.}
    \label{fig:journal}
\end{figure}

A plain journal bearing is shown in Figure \ref{fig:journal},  having a rotating inner shaft of radius $r$, with angular velocity $\omega$, and an axial offset $a$ from a stationary containing sleeve of radius $R$. Here we consider the simplest problem of a single fluid phase between the shaft and sleeve.

\subsubsection{No-slip lubrication analysis}
In this subsection, we overview the key assumptions and results of the standard no-slip lubrication analysis, closely following the exposition in \cite{sherman_viscous_1990}.

The radial clearance of the journal bearing is defined as $C=R-r$, and it is assumed that $C\ll R$ such that the curvature of the streamlines can be neglected and the local clearance is well approximated by:
\begin{equation}
H(\theta)=C(1- \eta \cos \theta )\, ,
\end{equation}
where $\eta=a/C$  is the shaft eccentricity (see Figure \ref{fig:journal}).
The no-slip solution predicts the fluid shear stress acting on the shaft,
\begin{equation}
    \tau_{0, \mathrm{shaft}}=\frac{2 \mu  R \omega  \left(2 \left(\eta ^2+2\right) \eta  \cos \theta -5 \eta ^2-1\right)}{C \left(\eta ^2+2\right) (\eta  \cos \theta -1)^2},
\end{equation} and the sleeve
\begin{equation}
    \tau_{0, \mathrm{sleeve}}=\frac{2 \mu  R \omega  \left(\left(\eta ^2+2\right) \eta  \cos \theta -4 \eta ^2+1\right)}{C \left(\eta ^2+2\right) (\eta  \cos \theta -1)^2}\, .
\end{equation} 
The resistive torque (per unit length) acting on the shaft is obtained by integration:
\begin{equation}
    T_0=-R^2 \int^{2\pi}_0  \tau_{0,\mathrm{shaft}} \, d \theta=\frac{4 \pi  \left(2 \eta ^2+1\right) \mu  R^3 \omega }{C \sqrt{1-\eta ^2} \left(\eta ^2+2\right)}\, . \end{equation}
Note,  since $C\ll R$ in this lubrication analysis, $r\approx R$.

Another important moment of the traction force for the journal bearing is the lift per unit length generated on the shaft ($\Lambda$); i.e., the net force in the direction of  $\theta=\pi/2$ (see Figure \ref{fig:journal}):
\begin{equation}
    \Lambda_0=\frac{12 \pi  \eta  \mu  R^3 \omega }{C^2 \sqrt{1-\eta ^2} \left(\eta ^2+2\right)} \, ,   \end{equation}
(note, there is a factor of $R$ missing in \cite{sherman_viscous_1990}).

A figure of merit for the journal bearing is given by a dimensionless ratio of the lift to the torque ($\mathcal{M}=R \Lambda/T$), providing a measure of the cost of producing lift. The figure of merit in no-slip conditions is
\begin{equation}
    \mathcal{M}_0=\frac{R \Lambda_0}{T_0}=\frac{3 R \eta  }{ C(1+ 2\eta ^2)} \, ,
\end{equation}
which shows that the greater the eccentricity ($\eta$) the greater the efficiency of the bearing (by this measure).

\subsubsection{First-order slip corrections}
Slip flow in journal bearings has been studied in a variety of contexts \citep{singh_effect_1984,shahdhaar_numerical_2020,arif_analysis_2022}, and is normally modelled using a modified Reynolds equation; i.e. using a lubrication analysis similar to the above, but with slip flow \citep{li_partially_2006, zhang_performance_2011,shahdhaar_numerical_2020,arif_analysis_2022}. To the author's knowledge, no \emph{analytical} solution to the slip-modified lubrication model has been presented for the journal bearing.

The first-order slip-correction coefficient to the retarding torque (which is the resistive moment) is obtained directly from Eq (\ref{resCorrFac}) (with $\mathcal{R}= T$, $L=C(1-\eta)$,  $\psi=1$, and $k = \omega$):
\begin{equation}\label{eq:jbT1}
    \hat{T}_1=-\frac{C(1-\eta)}{T_0 \omega \mu } \int_0^{2\pi}  (\tau_{0, \mathrm{shaft}}^2+\tau_{0, \mathrm{sleeve}}^2 )R\, \mathrm{d}\theta =-\frac{-8 \eta ^4+22 \eta ^2+4}{(\eta +1) \left(\eta ^2+2\right) \left(2 \eta ^2+1\right)}\, ,
\end{equation}
where the minimum clearance, $C(1-\eta)$, is taken as the characteristic scale of the bearing. Note, integration is over both surfaces, not just the shaft. This analytical solution for low-slip flow (which is equivalent to `slip-flow' conditions in gas bearings), tells us that slip will reduce the retarding torque for all values of eccentricity ($0\le\eta<1$).

The first-order slip correction for the lift force (which is not the resistive moment) requires evaluation of the more general expression for the slip-correction coefficient (\ref{firstOrderSCFmain}), and requires a conjugate solution: a no-slip solution to shaft translation, in a direction that we wish to evaluate the lift (in the direction $\theta=\pi/2$). The shear stress on both shaft and sleeve from a unit translational velocity of the shaft in a direction parallel to $\theta=\pi/2$ is:
\begin{equation}
   \tau_0'= -\frac{6 \mu  R \left(\left(\eta ^2+2\right) \cos \theta -3 \eta \right)}{C^2 \left(\eta ^2+2\right) (\eta  \cos \theta -1)^2}\, .
\end{equation}
Substitution into equation (\ref{firstOrderSCFmain}) (with $M= \Lambda$, $L=C(1-\eta)$ and  $\psi=1$) gives
\begin{equation}\label{eq:jbL1}
    \hat{\Lambda}_1=\frac{C(1-\eta)}{\Lambda_0 \mu } \int_0^{2\pi}  \tau'_0(\tau_{0, \mathrm{shaft}}+\tau_{0, \mathrm{sleeve}} ) R\, \mathrm{d}\theta =\frac{6 \left(\eta ^2-2\right)}{(\eta +1) \left(\eta ^2+2\right)}\, .
\end{equation}
Quick inspection of (\ref{eq:jbL1}) reveals that, as for retarding torque, the slip-correction coefficient for lift is negative for all values of eccentricity ($0\le\eta<1$). In other words, small amounts of slip will always reduce the lift generated by the journal bearing.

The figure of merit can be expanded as follows:
\begin{equation}
\hat{\mathcal{M}}=\frac{\hat{\Lambda}}{\hat{T}}=\frac{1+\hat{\Lambda_1}\xi+\mathcal{O}(\xi^2)}{1+\hat{T}_1\xi+\mathcal{O}(\xi^2)}=1+ \hat{\mathcal{M}}_1 \xi + \mathcal{O}(\xi^2) \, ,
\end{equation}
where 
\begin{equation}\label{eq:jbM1}
    \hat{\mathcal{M}}_1=\hat{L}_1-\hat{T}_1=\frac{4 (\eta -1)}{2 \eta ^2+1}\, .
\end{equation}
The immediate observation is that (\ref{eq:jbM1}) is necessarily negative: small amounts of slip will always lower the bearing's figure of merit. In other words, for low-slip flows, slip reduces the lift force proportionally more than it reduces the resistive torque. However, the impact of slip's negative effect on the figure of merit is reduced for greater eccentricities.

\subsubsection{Numerical verification}
The governing Reynolds equation for the fluid pressure in the bearing, assuming a uniform slip length $\ell$ on both shaft and sleeve, is:
\begin{equation}
   \frac{d}{d\theta} \left( H^2(H+6\ell) \frac{dp}{d\theta} \right)=6 \mu \omega R^2 \frac{d H}{d \theta} \, ,
\end{equation}
where $p$ is the pressure, and which upon integration gives:
\begin{equation}\label{JBdpdthetaNum}
  \frac{dp}{d\theta}=6 \mu \omega R^2 \frac{ H}{H^2(H+6\ell)}+\frac{A}{H^2(H+6\ell)} \, .
\end{equation}
The constant of integration, $A$, can be found by numerically integrating (\ref{JBdpdthetaNum}) with the condition that the pressure be continuous ($\int_0^{2\pi} \tfrac{dp}{d\theta} d\theta =0$). Subsequent integration of (\ref{JBdpdthetaNum}) provides the pressure distribution in the bearing.
Along with the shear-stress distribution on the shaft,
\begin{equation}
    \tau=-\frac{H}{2R}\frac{dp}{d\theta}-\frac{\mu \omega R}{2\ell+H}\, ,
\end{equation}
the lift and retarding torque on the shaft can be obtained.

\begin{figure}
    \centering
    \includegraphics[width=.9\linewidth]{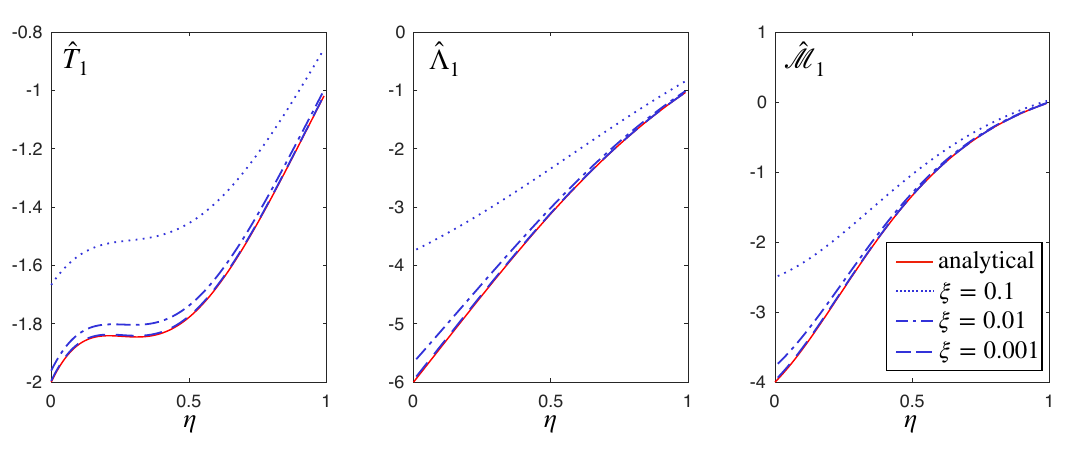}
    \caption{Comparison of numerical and analytical predictions for the first-order slip-correction coefficients. Analytical expressions for $\hat{T}_1$, $\hat{\Lambda}_1$ and $\hat{\mathcal{M}}_1$ (solid red lines) from Eqs (\ref{eq:jbT1}), (\ref{eq:jbL1}) and (\ref{eq:jbM1}), respectively. Numerical results for varying levels of slip: $\xi=l/(C(1-\eta))=0.1, 0.01, 0.001$.}
    \label{fig:journalResults}
\end{figure}

For each of the moments, $T$ and $\Lambda$, and the figure of merit $\mathcal{M}$, the first-order slip-correction coefficients are estimated from numerical calculations by: 
\begin{equation}
    \hat{M}_1\approx \frac{M-M_0}{M_0 \xi}\, ,
\end{equation}
where $\xi=l/L$ and $L=C(1-\eta)$. Figure \ref{fig:journalResults} compares these numerical results with the analytical results derived above; as expected, as the slip length is reduced, they converge. For larger slip lengths (relative to the minimum clearance), the numerical results differ from the analytical results, but the qualitative variation with changing eccentricity remains similar.

\subsection{Spherical squirmer}\label{sec:squirmer}

The squirmer, first proposed by \cite{lighthill_squirming_1952}, is a standard model for a self-propelled particle in Stokes flow. The spherical squirmer (of radius $R$) creates axi-symmetric surface motions, modelled by tangential and radial surface velocities ($u_\theta$, $u_r$), that in turn generate a translational axial velocity ($W$); see Figure \ref{fig:sphere} for the spherical coordinate system.

\subsubsection{The no-slip solution}
Lighthill first developed the no-slip analytical solution for the squirmer, but this was later corrected by \cite{blake_spherical_1971}, which we reproduce here in a slightly different form.

In the laboratory frame, the no-slip Stokes solution around the particle can be decomposed into a part due to translational wall motion ($\bo{u}_0^\mathrm{t}$, $\bo{\sigma}_0^\mathrm{t}$)  and a part due to wall motion relative to that translation, i.e. the squirming motion ($\bo{u}_0^\mathrm{s}$, $\bo{\sigma}_0^\mathrm{s}$):
\begin{equation}
\bo{u}_0=\bo{u}_0^\mathrm{t} + \bo{u}_0^\mathrm{s} \, ,   \qquad \bo{\sigma}_0=\bo{\sigma}_0^\mathrm{t} + \bo{\sigma}_0^\mathrm{s}
\end{equation}
where $\bo{u}_0$ and $\bo{\sigma}_0$ are the total velocity and stress fields of the no-slip solution, respectively. Both translational (superscript  t) and squirming (superscript s) components of the solution decay to zero in the far-field. The respective no-slip boundary conditions at the particle surface ($S_1$), in spherical polar coordinates, are:
\begin{equation}
\left. \begin{array}{ll}
                  {u}^\mathrm{t}_r(r,\theta)= W_0 \cos \theta  \mathrm{,} \quad     &
{u}^\mathrm{s}_r(r,\theta) =\sum
_{n=1}^{\infty} A_n P_n(\cos \theta) \vspace{.1cm}\\ 
              {u}^\mathrm{t}_\theta(r,\theta)= -W_0 \sin \theta  \mathrm{,} \quad  & {u}^\mathrm{s}_\theta(r,\theta) = \sum
_{n=1}^{\infty} B_n V_n(\cos \theta) 
                \end{array}  \right\} 
\quad \mathrm{for} \quad r=R
\end{equation}
where $r$ is the radial coordinate, $\theta$ is the polar angle, $W_0$ is the speed of particle translation along the polar axis ($\theta=0$), $A_n$ and $B_n$ are coefficients describing the form and strength of the squirming motion, $P_n$ are Legendre polynomials, and $V_n(\cos\theta)=-2/(n(n+1))\ dP_n(\cos\theta)/d\theta$. Note, here, we have restricted attention to volume-preserving wall motions (i.e. the particle cannot lose or gain mass).  

Blake's no-slip solution for the velocity field, decomposed into the two parts, is:
\begin{eqnarray}
{u^\mathrm{t}_r}&=& W_0 \cos (\theta ) \left(\frac{3 R}{2 r}-\frac{R^3}{2 r^3}\right) \, ,\\
{u^\mathrm{t}_\theta}&=&-W_0 \sin (\theta ) \left(\frac{R^3}{4 r^3}+\frac{3 R}{4 r}\right) \, , \\
{u^\mathrm{s}_r}&=& \frac{1}{2}\sum
_{n=1}^{\infty}  \left[ (n A_n -2 B_n)\frac{R^{n}}{r^{n}}+(2B_n-A_n(n-2))\frac{R^{n+2}}{r^{n+2}}\right] P_n(\cos \theta) \, , \\
{u^\mathrm{s}_\theta}&=&\frac{1}{4}\sum
_{n=1}^{\infty}  \Bigg[\left((4 -2 n) B_n + A_n n(n-2)\right)\frac{R^{n}}{r^{n}} \nonumber\\ && \hspace{3.4cm} +\left(2 n B_n-A_n n (n-2)\right)\frac{R^{n+2}}{r^{n+2}}\Bigg]  V_n(\cos \theta) \, ,
\end{eqnarray} 
which generates the following shear-stress ($\tau_{r\theta}$) components at the boundary: 
\begin{eqnarray} \label{eq:sqtau0t}
\tau^t_0&=& \frac{3 \mu  W_0 }{2 R} \sin (\theta ) \, ,
\\  \label{eq:sqtau0s}
\tau^s_0&=& -\frac{\mu}{R} \sum
_{n=1}^{\infty}\left(\tfrac{3}{2}  n A_n+ (2n+1)B_n\right) V_n(\cos \theta)  \, .
\end{eqnarray}
The motile force (of the fluid on the particle) generated by the squirming motion is obtained by integrating the induced traction force over the squirmer surface ($S_1$):
\begin{equation}\label{eq:sqf0}
F_0=
 \bo{i}_z \cdot \int_{S_1} \bo{\sigma}^s_0 \cdot \bo{n}\,\,dS= 2\pi \mu R  ( 2B_1-A_1)    ,
\end{equation}
which must be balanced, if the particle is self-propelled and there is no external force, by the drag generated in translation:
\begin{equation}
\label{eq:sqd0}
D_0=
 - \bo{i}_z \cdot \int_{S_1} \bo{\sigma}^t_0 \cdot \bo{n}\,\,dS =6\pi \mu R W_0  \, ,
\end{equation}
where $\bo{i}_z$ is a unit vector along the polar axis. Equating (\ref{eq:sqf0}) and (\ref{eq:sqd0}) leads to an expression for the translational speed of the particle:
\begin{equation}
W_0=\tfrac{1}{3}(2B_1-A_1) \, ,
\end{equation}
as obtained by Lighthill and Blake. The most significant implication of this result is that it is only the first modes of the radial and tangential surface motions that contribute to particle translation. 

\subsubsection{First-order slip corrections}

Here we consider the impact of a uniform slip length at the interface between the surface of the squirmer and the suspending fluid.\footnote{In some articles, the word `slip' is used to refer to the tangential motion of the squirmer's surface itself --- this is not what is meant here.} 

Similarly to previous sections, the motile force generated by the squirming motion is expanded to first order in slip length:
 \begin{equation}\label{eq:sqForce}
\frac{F}{F_0}=1+\hat{F}_1 \,\xi+ \mathcal{O}\left(\xi^2\right)  
\end{equation}  
where $\xi=l/R$, and $\hat{F}_1=F_1/F_0$ is the first-order slip-correction coefficient that we wish to find. To obtain it, we need the general expression for the first-order slip-correction coefficient, (\ref{firstOrderSCFmain}), and the no-slip shear-stress distribution associated with the squirming motion, $\bo{\tau}_0^s$, from equation (\ref{eq:sqtau0s}). Additionally, for the conjugate Stokes flow, we need the no-slip solution for unit translation ($\bo{\tau}_0'=\bo{\tau}_0^t/W_0$), 
so that the moment of the traction force obtained is in the direction of translation. From Eq. (\ref{firstOrderSCFmain}), with $M_i=F_i$, $L=R$ and $\psi=1$, we obtain: \begin{equation}\label{eq:sqf1}
\hat{F}_1= \tfrac{R}{F_0 \mu W_0} \int_{S_1}  \, \bo{\tau}^\mathrm{t}_0\cdot\bo{\tau}^\mathrm{s}_0\, \, dS= \tfrac{2\pi R^3}{ F_0 \mu W_0}  \int^{\pi}_0 \, {\tau}^\mathrm{t}_0 \,{\tau}^\mathrm{s}_0\, \, \sin{\theta} \,d\theta=\frac{3(A_1+2B_1)}{A_1-2B_1} \, .
\end{equation}
From \S\ref{sec:simpleSphere}, the slip drag on a translating sphere is shown to be:
\begin{equation}\label{eq:sqDrag}
\frac{D}{D_0}=1- \xi+\mathcal{O}\left(\xi^2\right)\,.
\end{equation}
Now, combining Equations (\ref{eq:sqd0}), (\ref{eq:sqf1}) and (\ref{eq:sqDrag}) with the condition for self-propulsion ($F=D$), an expression for the translational velocity is found:
\begin{equation}\label{WslipCorr}
  \frac{W}{W_0} =1+4\left(\frac{A_1+B_1}{A_1-2 B_1}\right)\xi +\mathcal{O}(\xi)^2  \, .
\end{equation}
This tells us some interesting things about the impact of low levels of slip on the squirmer's swimming speed. For purely tangential squirming motion at the surface ($A_1=0$), slip hinders swimming ($W/W_0<1$). This is because the motile force generated by pure tangential motion is reduced by slip at three times the rate of the translational drag. Conversely, for purely radial wall motion ($B_1=0$), slip promotes swimming speed ($W/W_0>1$), because, in this case, motile force is actually increased by low-levels of slip.  

Table \ref{table:sqW} provides numerical verification of equation (\ref{WslipCorr}) for three different cases; see Appendix \ref{sec:MFS} for numerical details. 

\begin{table}
 \begin{spacing}{1.15}
\centering
\begin{tabular}{c |c c c | c}
\hline \hline 
& \multicolumn{3}{c|}{Numerical}  &   Eq.\,(\ref{WslipCorr})
\\
  &  $\xi=10^{-2}$ & $\xi=10^{-3}$ & $\xi=10^{-4}$& $\xi\to0$   \\
\hline
$A_1=0 $\,\,\,\, & $-1.9608$ &$-1.9960$  &$-1.9996$& $-2$\\
 $B_1=0$\,\,\,\, & \,\, 3.9216 & \,\, 3.9920 &\,\, 3.9989 & \,\, 4 \\
$A_1=B_1 $ & $-7.8431$ &$-7.9840$  &$-7.9980$ & $-8$\\
\hline
\end{tabular}
\end{spacing}
\caption{Comparison of numerical and analytical results: $(W/W_0-1)/\xi$}
 \label{table:sqW}
\end{table}

\subsection{Poiseuille flow through arbitrary cross-section channels}\label{sec:channel}
\begin{figure}
    \centering
    \includegraphics[width=0.55\linewidth]{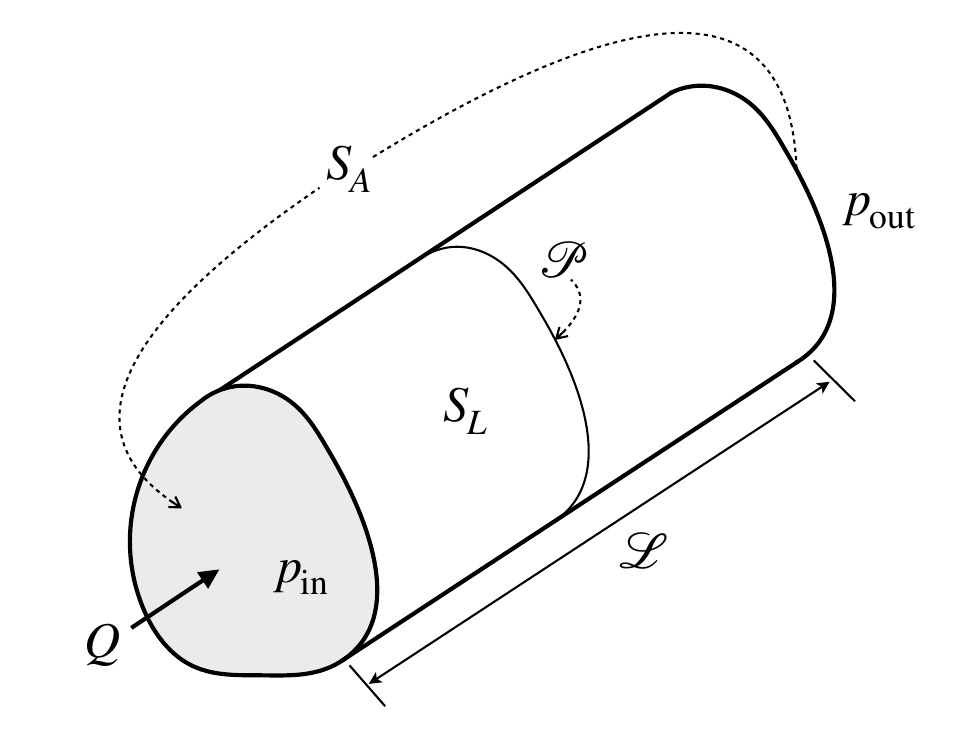}
    \caption{A channel flow with arbitrary cross-section and an applied pressure drop ($\Delta p=p_\mathrm{in}-p_\mathrm{out}$).}
    \label{fig:channel}
\end{figure}

In this final example, we consider an internal flow containing inflow and outflow boundaries. Figure \ref{fig:channel} shows a long straight channel (length $\mathcal{L}$) with an arbitrary, but constant, cross-section of area $A$. The pressure gradient is assumed constant throughout the channel,  $\bo{\nabla} p=\bo{i}(p_\mathrm{out}-p_\mathrm{in})/\mathcal{L}$, where  $\bo{i}$ is a unit vector along the channel length, which generates a volumetric flow rate $Q$. The boundary of the fluid domain is separated into two parts: the walls of the channel ($S_L$), at which there is the potential for slip, which we assume to be constant in the streamwise direction, and the inlet and outlet boundaries ($S_A$) at which $\psi=0$.

Our aim in this section is to find the first-order impact of slip on the pressure drop ($\Delta p=p_\mathrm{in}-p_\mathrm{out}$)  for a given flow rate. A force balance in the direction of the channel gives us the  pressure drop in terms of a traction-force moment over the channel walls:
\begin{equation}
    \Delta p=\frac{1}{A}\int_{S_L} \bo{i}\cdot \bo{\sigma}\cdot \bo{n} \,\, dS \, ,
\end{equation}
which we expand as previously:
\begin{equation}
    {\Delta p}=    {\Delta p}_0 + {\Delta p}_1\xi + ... \, ,
    \end{equation}
where $\xi=l/L$ and the charactersitic length scale $L$ is chosen based on the cross-section.

To evaluate the first-order slip-correction coefficient for this  moment we need the general expresssion from \S\ref{sec:findingM1}, repeated here for convenience:
\begin{subequations}
\label{eq:M1repeated}
  \begin{align}
M_1& =
\int_{S} \bo{u}'_0\cdot \bo{\sigma_1}\cdot \bo{n} \,\, dS \, , \\
 &= \tfrac{L}{ \mu} \int_{S} \psi \, \bo{\tau}_{0}\cdot\bo{\tau}'_{0}\, \, dS\, .
 \end{align}
\end{subequations} 
 The conjugate solution to obtain the deisired moment is a simple transformation of the original no-slip solution:
 \begin{subequations}\label{eq:conjChannel}
\begin{equation}
    \bo{u}_0'=-\frac{{\bo{u}_0}}{Q}+\frac{\bo{i}}{A}\, \quad \mathrm{and} \quad \bo{\sigma}'_0=-\frac{\bo{\sigma}_0}{Q} \, . \tag{\ref{eq:conjChannel} \emph{a,b}}
\end{equation}
\end{subequations}
To demonstrate this choice gives the correct moment, we substitute the conjugate velocity field (\ref{eq:conjChannel}$a$) into
(\ref{eq:M1repeated}$a$) (noting that $\int_{S_A} \bo{u}_0' \, dA =\bo{0}$ and $\bo{u}_0=\bo{0}$ at $S_L$): 
\begin{eqnarray}\label{eq:chint1}
M_1
=\frac{1}{A}\int_{S_L} \bo{i}\cdot \bo{\sigma_1}\cdot \bo{n} \,\, dS -\Delta p_1 \bo{i}\cdot \int_{S_A}  \bo{u}'_0 \,\, dA 
=\Delta p_1 \, .
\end{eqnarray}

Finally, substituting the conjugate stress field (\ref{eq:conjChannel}$b$) into (\ref{eq:M1repeated}$b$), and recalling $\psi=0$ at $S_A$, gives:
\begin{eqnarray}\label{eq:chFOCFmain}
{\Delta p}_1 &=&- \tfrac{L \mathcal{L}}{Q \mu} \int_{\mathcal{P}} \psi \, {\tau}_{0}^2\, \, d\mathcal{P} \, . 
\end{eqnarray}

As is intuitive, perhaps, the first-order slip correction is negative for any distribution of positive slip length around the perimeter, $\mathcal{P}$, of any cross-sectional shape: low levels of slip will always reduce the pressure loss in a Poiseuille flow for a given flow rate.

\subsubsection{Hagen-Poiseuille Flow}
If the channel is a circular cross-section of radius $R$, the no-slip solution for the pressure drop is the familiar Hagen-Poiseuille equation:
\begin{equation}
  {\Delta p}_0 =\tfrac{8 \mu \mathcal{L} Q}{\pi R^4} \, ,  
\end{equation}
and the corresponding shear-stress magnitude is:
\begin{equation}
  {\tau}_0 =\frac{\Delta p_0 R}{2\mathcal{L}}\, .
\end{equation}
The first-order slip-correction coefficient from (\ref{eq:chFOCFmain}), with $L=R$, is simply:
\begin{eqnarray}\label{eq:chhpfocf1}
\hat{{\Delta p}}_1 &=&- 4 \bar{\psi} , 
\end{eqnarray}
where $l \bar{\psi}$ is the average slip length over the channel perimeter.  This tell us, in the example of a circular channel coated with heterogenous regions of equal and constant slip, the first-order slip-correction coefficient is directly proportional to the area of the surface coating (the specific distribution is unimportant). 

The full-slip solution to the Hagen-Poiseuille equation for constant slip is 
\begin{equation}
   \hat{\Delta p}= 1/(1+4\xi) \, ,
\end{equation}
where $\xi=l/R$,  and which, expanded to the first-order, is:
\begin{equation}
   \hat{\Delta p}= 1-4\xi +\mathcal{O}(\xi^2) \, ,
\end{equation}
agreeing with the first-order slip-correction coefficient obtained in (\ref{eq:chhpfocf1}) for $\bar{\psi}=1$.

\section{Rate of work and energy dissipation}\label{sec:energy}
    For certain applications, there is interest in the impact of slip on the power expenditure of a bounding surface. In this section we start by evaluating the first-order slip-correction coefficient for power expenditure (the rate of work done on the fluid) and next ask the question: where is the energy dissipated in low-slip flows?

\subsection{First-order slip-correction for rate of work}
The rate of work done by a bounding surface, $S$, on the fluid is given by:
   \begin{eqnarray}\label{power}
P=-\int_{S} \bo{U} \cdot\bo{\sigma} \cdot  \bo{n} \, \, dS  \, ,
\end{eqnarray}
where, as a reminder, $\bo{n}$ is a surface-normal facing into the fluid; note, in the absence of external forces, $P\ge0$. Substituting (\ref{stressSeries}) into (\ref{power}), and recalling that $\bo{u_0}=\bo{U}$, gives:  
 \begin{eqnarray}
P=P_0 +\xi P_1 + \, ... \, ,
\end{eqnarray}
 where
 \begin{equation}\label{eq:pP0}
P_0=-\int_{S} \bo{u}_0 \cdot\bo{\sigma}_0 \cdot  \bo{n} \, \, dS \, ,
\end{equation} and (again, assuming no external forces)
\begin{equation}\label{eq:pP1}
P_1=- \int_{S} \bo{u}_0 \cdot\bo{\sigma}_1 \cdot  \bo{n} \, \, dS \, = - \int_{S} \bo{u}_1 \cdot\bo{\sigma}_0 \cdot  \bo{n} \, \, dS \, .
\end{equation}
Upon substituting the first-order solution for velocity, $\bo{u}_1$,  from equation (\ref{bcSeries}),  we obtain an expression for the first-order slip-correction coefficient for $P$: 
\begin{equation}\label{eq:powerfocf}
\hat{P}_1= -\frac{ L}{P_0 \mu } \int_{S} \psi \, \tau_0^2\, \, dS \, .
\end{equation}
Combining (\ref{eq:rmR0}) , (\ref{resCorrFac}), (\ref{eq:pP0}) and (\ref{eq:powerfocf}) shows that:
\begin{eqnarray}\label{powerid1}
    \hat{P}_1 \equiv \hat{\mathcal{R}}_1 \, .
\end{eqnarray}
In other words, the first-order slip-correction coefficient for the resistive moment is equivalent to the first-order slip-correction coefficient for the rate of work done by the boundary on the fluid ($P$).  Consequently, for every result derived in the paper for the resistive moment (eqns (\ref{eq:sprm1}), (\ref{eq:sprm2}), (\ref{eq:resMomVarSphere}), (\ref{eq:restorvarsph}), (\ref{eq:prosc}), (\ref{eq:oblsc}) and (\ref{eq:jbT1})) we have also obtained $\hat{P}_1$. 

In general, then, we can say that a small amount of slip will not only reduce the resistive moment for the flow in question (as discussed in \S\ref{sec:resmom}), but also the work done by the bounding surface on the fluid volume by the same proportion.

\subsection{Energy dissipation}
In no-slip conditions, and in the absence of body forces, the rate of energy dissipation in the fluid volume ($\Phi$) is equal to the rate of work done by the bounding surface on the fluid. Interestingly, this is not the case when there is slip; $P\ne\Phi$. The energy dissipation in the fluid volume is given by:
   \begin{eqnarray}\label{eq:energydiss}
\Phi=-\int_{S} \bo{u} \cdot\bo{\sigma} \cdot  \bo{n} \, \, dS\, .
\end{eqnarray}
Substituting (\ref{uSeries}) and (\ref{stressSeries}) into (\ref{eq:energydiss}) gives:  
 \begin{eqnarray}
\Phi=-\int_{S} \bo{u}_0 \cdot\bo{\sigma}_0 \cdot  \bo{n} \, \, dS - \xi\int_{S} \bo{u}_1 \cdot\bo{\sigma}_0 \cdot  \bo{n} \, \, dS  -\xi \int_{S} \bo{u}_0 \cdot\bo{\sigma}_1 \cdot  \bo{n} \, \, dS  \, + \, ...
\end{eqnarray}
Following similar steps to before, this simplifies to
 \begin{eqnarray}
\hat{\Phi}= 1+ \hat{\Phi}_1 \xi + \mathcal{O}(\xi^2)
\end{eqnarray}
where $\hat{\Phi}_1=\Phi_1/\Phi_0$, $\Phi_0=P_0$ and 
\begin{eqnarray}
    \hat{\Phi}_1 \equiv 2\, \hat{\mathcal{R}}_1 \, .
\end{eqnarray}
The difference between the rate of work done on the fluid and the rate of energy dissipated in the fluid is due to energy dissipation at the slip interface itself, $E$. If this is expanded in slip length we get:
\begin{equation}
\hat{E}= \hat{E}_1 \xi +\mathcal{O}(\xi)^2  \, ,
\end{equation}
where
\begin{equation}
 \hat{E}_1={\hat{P}_1-\hat{\Phi}_1}=-\hat{\mathcal{R}}_1 \, .
\end{equation} 
and where, since $\hat{\mathcal{R}}_1<0$, $\hat{E}_1>0$. This means that any slip-induced reduction in the rate of work done on the fluid by the bounding surface is attended by both a reduction in the rate of dissipation in the fluid and an {\emph{increase}} in the dissipation at the slip interface.

\section{Summary and Discussion}\label{sec:discussion}
A convenient method for deriving analytical solutions to Stokes flows with low levels of slip is presented, relevant to applications where the slip length is small compared to the geometry. In general, these first-order approximations to slip Stokes flows are both much simpler to derive and much simpler to evaluate than the full slip solution. Of course, in many situations, the slip length will not be small, and the methods presented here can only be considered approximate. For example, for the drag on a sphere (of radius $R$) with slip length ($l$), the percentage error of the first-order approximation is given by:
\begin{equation}
    \mathrm{\%\,error}=100\times \frac{3 \xi^2}{2\xi+1}\, ,
\end{equation}
where $\xi=l/R$. The error increases from 2.5\% at $\xi=0.1$ to nearly 40\% at $\xi=0.5$, to greater than 100\% for $\xi>1$. Note, though, for rarefied-gas `slip flows', the first-order approximation is, in fact, the only valid one.  

Numerically calculating first-order slip-correction factors from the gradient of the property in question (e.g. drag) can be extremely computationally demanding. This is because its evaluation requires calculating the difference between a very low-slip solution and the no-slip solution --- a difference that is very small, and thus hard to calculate accurately. The numerical techniques used in this paper to verify the derived analytical expressions (the Method of Fundamental Solutions (MFS), see Appendix A) are very high accuracy for certain classes of geometry. However, in general, the gradient approach to calculating the slip-correction coefficients is not straightforward. 

The expressions derived in \S\ref{sec:theory} offer a more convenient numerical method of obtaining the first-order slip-correction coefficients: one involving numerical integration of the square of the shear-stress magnitude from an analytical/numerical no-slip solution to the flow problem(s).  What is required, is an accurate means of integrating properties over the bounding surfaces --- in the case of an MFS framework a convenient solution exists for closed surfaces \citep{lockerby_integration_2022}, but a number of techniques can be employed.

An alternative is to calculate  $\bo{u}_1, \bo{\sigma}_1$ directly, by solving the Stokes equations with the boundary conditions taken from the no-slip stress, $\bo{\sigma}_0$, as per equation (\ref{bcSeries}). This allows the construction of the whole slip solution, to first order, using the original expansion: (\ref{uSeries}) and (\ref{stressSeries}). In a similar way, higher-order solutions, and their associated traction-force moments, could also be obtained, allowing better predictions at higher $\xi$. Of course, at some value of $\xi$, the series will diverge, and so additional terms in the expansion will yield diminishing returns.

In this article, some general results pertaining to Stokes flow in low-slip conditions have been derived. For example, consider the rotation of a particle of arbitrary geometry driven by an external torque. The addition of any (small) slip length, however it is distributed across the particle surface, will reduce the torque required to maintain the particle's rotational speed. This is because, in general, the correction coefficient for the resistive moment, equation (\ref{resCorrFac}), is always negative (for positive slip lengths).
In addition, it was shown in \S\ref{sec:energy} that, keeping with the same example, the rate in which the torque is reduced by slip is equal to the rate in which slip reduces the work done by the particle on the fluid. In this context, the result is obvious, since power to drive the particle at a fixed angular speed is proportional to the retarding torque. What is more surprising is that the rate in which energy dissipates in the fluid reduces at \emph{twice} the rate that the applied power reduces. The missing energy must be dissipated at the fluid-solid interface. Whether this exists in real conditions, or even in molecular simulations, is unclear; but it is reasonable to expect strong viscous dissipation in an interfacial layer where there are strong velocity gradients. 

\subsubsection*{Acknowledgments} The author would like to thank James Sprittles for very helpful comments on a draft of the manuscript.

\subsubsection*{Funding}This work was supported by the EPSRC under grant EP/V01207X/1.

\subsubsection*{Declaration of interests}The author reports no conflict of interest.

\subsubsection*{Data availability statement}The data that support the findings of this study are tabulated within the article.

\subsubsection*{Author ORCID} D. A. Lockerby, https://orcid.org/0000-0001-5232-7986

\appendix
\section{The Method of Fundamental Solutions}\label{sec:MFS}

For the numerical calculation of the external Stokes flows considered in \S\ref{sec:simpleSphere}, \S \ref{sec:variableSurfaceSphere}, \S \ref{sec:spheroids} and \S\ref{sec:squirmer},  we employ the Method of Fundamental Solutions (MFS) \citep{lockerby_fundamental_2016, cheng_overview_2020}. The MFS uses a superposition of fundamental solutions to the Stokes equations (popularly known as Stokeslets) to construct an analytical solution that approximately satisfies the boundary conditions at the particle surface (a zero disturbance far-field condition is automatically satisfied by the Stokeslets). The numerical procedure has much in common with the Boundary Element Method, requiring the surface of the particle to be discretised as opposed to the volume of fluid it occupies; this reduces the dimensionality of the numerical calculation and removes the requirement for a finite fluid domain.  The primary numerical parameter is the number of `boundary nodes' (sometimes referred to as collocation points) that discretise the particle boundary, $N$; as $N$ increases the superposed solution becomes a more accurate representation of the true one.

\begin{figure}
    \centering
    \includegraphics[width=0.72\linewidth]{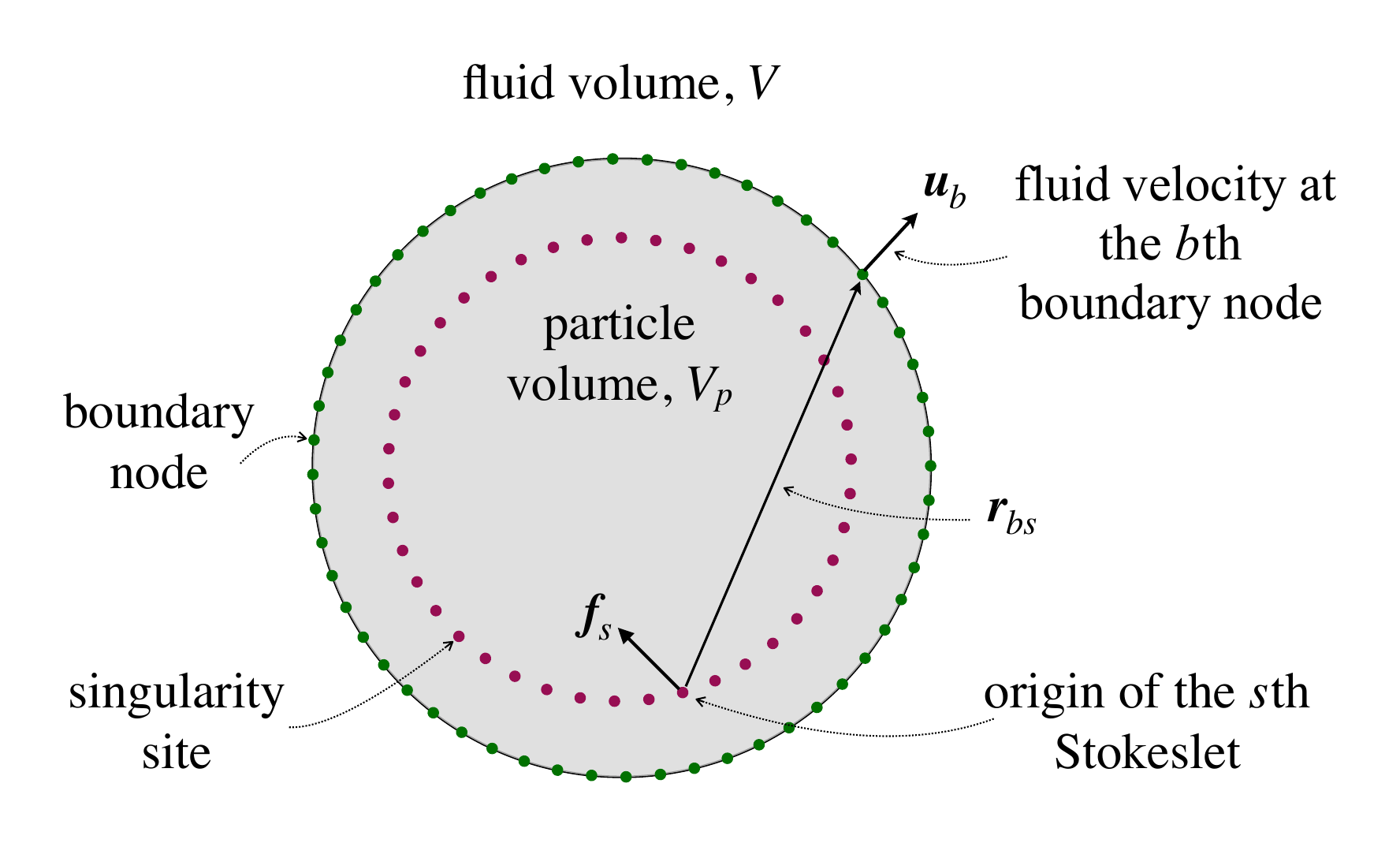}
    \caption{Illustration of site and node arrangement in the Method of Fundamental Solutions applied to external flows around particles.}
    \label{fig:MFS}
\end{figure}

The Stokeslet can be viewed as the Stokes-flow response to a steady-state point forcing in 3D space. The MFS approximates a given flow using a superposition of these Stokeslets, with the location of the point forces (referred to as singularity sites) distributed \emph{outside} of the fluid domain (e.g. set within the volume of a solid particle, $V_p$, see Figure \ref{fig:MFS}). 

The governing equations that are solved (exactly) by the MFS are
\begin{equation} \label{eq:stokesMFS}
  \bo{\nabla} \cdot \bo{u}=\bo{0} \, , \quad   \bo{\nabla} \cdot 
 \bo{\sigma} =-\sum\limits_{s=1}^{M} \bo{f}_s   \delta(\bo{r}-\bo{r}_s)   \, ,
\end{equation}
where $\bo{f}_s$ is the force (located at $\bo{r}_s$) associated with the $s$th Stokeslet, and $\delta$ is the Dirac delta function. Note, these revert to the Stokes equations, (\ref{eq:contmom}), within the fluid volume (assuming no external forces). The corresponding analytical solution is:
\begin{equation} \label{eq:MFSuAn}
\bo{u}=\frac{1}{8 \pi \mu} \sum\limits_{s=1}^{M}\bo{f}_s\cdot \left( \frac{\mathbb{I}}{\|\bo{r}-\bo{r}_s\|} + \frac{(\bo{r}-\bo{r}_s)(\bo{r}-\bo{r}_s)}{\|\bo{r}-\bo{r}_s\|^3} \right)
  \end{equation}
     \begin{equation} \label{eq:MFSsAn}
\bo{\sigma}=-\frac{3  }{4 \pi } \sum\limits_{s=1}^{M}\bo{f}_s \cdot\left(\frac{(\bo{r}-\bo{r}_s)(\bo{r}-\bo{r}_s)(\bo{r}-\bo{r}_s)}{\|\bo{r}-\bo{r}_s\|^5} \right)
  \end{equation}
where $\mathbb{I}$ is the identity tensor, and $\|\cdot\|$ denotes the Euclidean (L2) norm. The primary aim of the MFS is to determine the Stokeslet forces, $\bo{f}_s$, from the boundary conditions of the problem.

Evaluating the velocity and stress field at the boundary nodes gives: 
  \begin{equation} \label{eq:appUb}
   \bo{u}_{b}=\frac{1}{\mu}\sum\limits_{s=1}^{M}   \bo{f}_s \cdot \mathbb{J}_{\,bs} 
\quad
\mbox{and} \quad \bo{\sigma}_{b}=-\sum\limits_{s=1}^{M}  \bo{f}_s \cdot \mathbb{K}_{bs} \, ,
 \end{equation}
 where
 \begin{equation} 
\mathbb{J}_{bs}=  \frac{1}{8 \pi}\left( \frac{\mathbb{I}}{\|\bo{r}_{bs}\|} + \frac{\bo{r}_{bs} \bo{r}_{bs}}{\|{\bo{r}}_{bs}\|^3} \right)\, \quad \mathrm{and} 
\quad \mathbb{K}_{bs}=\frac{3}{4 \pi}\left( \frac{ {\bo{r}}_{bs} {\bo{r}}_{bs}{\bo{r}}_{bs} }{\|{\bo{r}}_{bs}\|^5} \right)\, ,
 \end{equation}
and where $\bo{r}_{bs}=\bo{r}_b-\bo{r}_s$, and $\bo{r}_b$ is the position of the $b$th boundary node; see Figure \ref{fig:MFS}. Substituting (\ref{eq:appUb}) into (\ref{eq:navierslip}) allows the Navier slip boundary condition at node $b$ to be written:
\begin{equation} 
\bo{U}_b=\frac{1}{\mu}\sum\limits_{s=1}^{M}\bo{f}_s\cdot \mathbb{L}_{bs} 
 \end{equation}
 where 
 \begin{equation} \label{bbL}
\mathbb{A}_{bs}= \mathbb{J}_{bs}+\ell_b  (\bo{n}_b \cdot   \mathbb{K}_{bs})\cdot (\mathbb{I} -\bo{n}_b \bo{n}_b )\, .
 \end{equation}
The rank-4 tensor, $\mathbb{A}_{bs}=A_{bsij}$, and rank-2 tensors, $\bo{f}_s=f_{sj}$ and $\bo{U}_b=U_{bi}$, can be reshaped with the bijection
\begin{equation}\nonumber
p=3(b-1)+i \,  \quad  \mathrm{and} \quad  q=3(s-1)+j\, , 
\end{equation}
 to obtain
\begin{equation}\label{newA2} 
\tilde{A}_{pq}=A_{bsij}, \, \quad \,  \tilde{f}_{q}=f_{sj}   \, , \quad \mathrm{and} \quad
\tilde{U}_{p}=U_{bi} 
\end{equation}
where $\tilde{\mathbb{A}}$ has size ($3N \times 3M$) , $\tilde{\bo{f}}$ has size ($3M\times 1$),  $\tilde{\bo{U}}$ has size ($ 3N \times 1$),  such that the evaluation of the Navier slip condition at all nodes is represented by the matrix equation:
\begin{equation}\label{eq:Axb}
\tilde{\mathbb{A}} \cdot \tilde{\bo{f}}= \tilde{\bo{U}} \, ,
\end{equation} 
which can be solved for the vector of force components ($\tilde{\bo{f}}$) by any standard linear-equation solver.  In practice, it benefits the numerics to have fewer Stokeslets than boundary nodes, which creates an overdetermined system that can be solved using a linear least-squares method. In this work, $M\approx 0.9 N$.

For the simulations presented in this article, the boundary nodes on the surface of each object (spheres or spheroids) are distributed evenly, and found using the Matlab code (DistMesh) written by \cite{persson_simple_2004}. In the MFS literature the singularity sites are often referred to as ‘source nodes’ and considerable work has been done on deciding how they should be optimally located \citep{karageorghis_practical_2009,chen_choosing_2016}. However, for the problems considered here, a simple surface-normal projection into the particle works well, i.e.: $\bo{r}_s= \bo{r}_b-\alpha_b \bo{n}_b$, where $\alpha_b$ is chosen such that each site is 5\% closer to its respective node than any other. After locating the sites in this way, approximately 10\% of them are deleted, so that $M\approx 0.9N$.

Once the forces are known, it is simple to calculate moments of the traction force on a particle via (\ref{eq:stokesMFS}) and the divergence theorem. For example, the net traction force $\bo{F}$ on the particle 
surface is simply a summation of the Stokeslet forces:
\begin{equation} \label{eq:consR}
\bo{F}=\int_{S} \bo{\sigma}\cdot \bo{n} \, dS=\int_{V_p} \bo{\nabla}\cdot \bo{\sigma} \, dV_p= - \int_{V_p} \sum\limits_{s=1}^{M} \bo{f}_s  \delta(\bo{r}-\bo{r}_s)\, dV_p = - \sum\limits_{s=1}^{M} \bo{f}_s \, .
\end{equation}

\subsection{Numerical convergence}\label{sec:MFSconvergence}
To verify the implementation of the MFS for Stokes slip flows, we compare numerical predictions of drag around a sphere with the classic analytic result due to \cite{basset_treatise_1888}, equation (\ref{eq:basset}). Table \ref{tab:MFSconvergence} shows convergence of the MFS result to the exact drag solution with increasing number of nodes ($N$) for a range of non-dimensional slip lengths.

\begin{table}
 \begin{spacing}{1.15}
    \centering
     \begin{tabular}{c|cccccc}
            \hline
         \hline 
          & \multicolumn{6}{c}{Non-dimensional slip length} \\
          &  $\xi$=0.1&  $\xi$=0.5&  $\xi$=1.0&  $\xi$=5.0&  $\xi$=10& \\
          \hline 
             $N$=24&  0.923702&  0.800570&  0.750400&  0.687590&  0.677452& \\
         $N$=32&  0.923193&  0.800117&  0.750082&  0.687515&  0.677421& \\
        $N$=66 &   0.923090&  0.800020&  0.750015&  0.687501&  0.677418& \\
$N$=156 & 0.923077&  0.800000&  0.750000& 0.687500& 0.677419&
        \\
        \hline 
Basset, Eq\,.(\ref{eq:basset}) & 0.923077&  0.800000&   0.750000&  0.687500&  0.677419& 
    \\
    \hline 
    \end{tabular}
    \end{spacing}
     \caption{Results for drag on a translating sphere in Stokes flow with slip (normalised with no-slip drag). Comparison of an analytical solution \citep{basset_treatise_1888} to the MFS.}    \label{tab:MFSconvergence}
\end{table}

Obtaining slip-correction coefficients from numerical simulations can be more demanding. For example, in Table \ref{table:pro}, the quantity of interest is $(\hat{D}-1)/\xi$. To obtain this quantity accurate to 4 significant figures, in the case when $\xi=10^{-5}$, requires the calculation of $\hat{D}$ accurate to 9  significant figures. The accuracy of the MFS is therefore essential for the purposes of verification.

Table \ref{table:NFScasesN} summarises the number of boundary nodes used in the various verification cases of the main article. Halving the number of nodes used in each case (or quartering, in the case of the oblate spheroid) results in a small change in the presented results; see the penultimate column of Table \ref{table:NFScasesN}. 

The MFS performs far worse when particles have sharp edges or high aspect ratio. In the example of the variable-slip-length sphere, \S\ref{sec:variableSurfaceSphere}, the MFS has to resolve discontinuities in slip length across the sphere's surface (see Figure \ref{fig:Janus2}). As such, this represents the most challenging case of the article, and requires a large number of boundary nodes to get accurate results.

\begin{table}
      \begin{spacing}{1.15}
    \centering
    \begin{tabular}{c|c|c|c|p{3.2cm}|c}   
    \hline \hline
Description &  {Table and section} & Case(s) & $N$ &  \centering Max \%$\Delta$ in tabulated results due to reducing $N$ by $X$ & $X$ \\
  \hline \hline
  Variable-slip sphere   &  Tables \ref{table:vs1} \& \ref{table:vs2},  \S\ref{sec:variableSurfaceSphere} & all & 3744 & \centering 0.5\% &  50\% \\ \hline
  \multirow{ 5}{*}{ Prolate Spheroids} & \multirow{ 5}{*}{Table \ref{table:pro}, \S \ref{sec:spheroids}}  & $E=1.1$ & 1376 & \centering  \multirow{ 5}{*}{0.02\%} & \multirow{ 5}{*}{50\%}\\
      &   &$E=1.5$ & 1744 &  & \\
    & &$E=2$  & 2160 &  & \\
    &  &$E=3$ & 3056 &  & \\
    &  &$E=4$ & 3872 &  & \\
\hline
  \multirow{ 5}{*}{ Oblate Spheroids} & \multirow{ 5}{*}{Table \ref{table:obl}, \S \ref{sec:spheroids}}  & $E=1.1$ & 540 & \centering  \multirow{ 5}{*}{0.02\%} & \multirow{ 5}{*}{25\%}\\
      &   &$E=1.5$ & 834 &  & \\
    & &$E=2$  & 1242 &  & \\
    &  &$E=3$ & 2376 &  & \\
    &  &$E=4$ & 3746 &  & \\ \hline
    Spherical squirmer & Table \ref{table:sqW}, \S\ref{sec:squirmer} & all & 1866 & \centering 0.05\% & 50\% \\ \hline
 \end{tabular}
         \end{spacing}
         \caption{Number of boundary nodes used in MFS verification simulations}
\label{table:NFScasesN}
\end{table}

\bibliographystyle{jfm}
\bibliography{references}

\begin{thebibliography}{38}
\expandafter\ifx\csname natexlab\endcsname\relax\def\natexlab#1{#1}\fi
\def\au#1{#1} \def\ed#1{#1} \def\yr#1{#1}\def\at#1{#1}\def\jt#1{\textit{#1}}
  \def\bt#1{#1}\def\bvol#1{\textbf{#1}} \def\vol#1{#1} \def\pg#1{#1}
  \def\publ#1{#1}\def\arxiv#1{#1}\def\org#1{#1}\def\st#1{\textit{#1}}

\bibitem[Arif {\em et~al.\/}(2022)Arif, Kango \& Shukla]{arif_analysis_2022}
{\sc \au{Arif, M.}, \au{Kango, S.} \& \au{Shukla, D.~K.}} \yr{2022}
  \at{Analysis of textured journal bearing with slip boundary condition and
  pseudoplastic lubricants}.  \jt{International Journal of Mechanical Sciences}
   \bvol{228},  \pg{107458}.

\bibitem[Arkilic {\em et~al.\/}(1997)Arkilic, Schmidt \&
  Breuer]{arkilic_gaseous_1997}
{\sc \au{Arkilic, E.B.}, \au{Schmidt, M.A.} \& \au{Breuer, K.S.}} \yr{1997}
  \at{Gaseous slip flow in long microchannels}.  \jt{Journal of
  Microelectromechanical Systems}  \bvol{6}~(2),  \pg{167--178}.

\bibitem[Basset(1888)]{basset_treatise_1888}
{\sc \au{Basset, A.B.}} \yr{1888} {\em A treatise on hydrodynamics: with
  numerous examples\/}, ,  \vol{vol.~2}.  \publ{Deighton, Bell and Company}.

\bibitem[Belyaev \& Vinogradova(2010)]{belyaev_effective_2010}
{\sc \au{Belyaev, A.~V.} \& \au{Vinogradova, O.~I.}} \yr{2010}  \at{Effective
  slip in pressure-driven flow past super-hydrophobic stripes}.  \jt{Journal of
  Fluid Mechanics}  \bvol{652},  \pg{489--499}.

\bibitem[Blake(1971)]{blake_spherical_1971}
{\sc \au{Blake, J.~R.}} \yr{1971}  \at{A spherical envelope approach to ciliary
  propulsion}.  \jt{Journal of Fluid Mechanics}  \bvol{46}~(1),  \pg{199--208}.

\bibitem[Cercignani(1969)]{cercignani_mathematical_1969}
{\sc \au{Cercignani, Carlo}} \yr{1969} {\em Mathematical {Methods} in {Kinetic}
  {Theory}\/}.  \publ{Boston, MA: Springer US}.

\bibitem[Chen {\em et~al.\/}(2016)Chen, Karageorghis \& Li]{chen_choosing_2016}
{\sc \au{Chen, C.~S.}, \au{Karageorghis, A.} \& \au{Li, Yan}} \yr{2016}  \at{On
  choosing the location of the sources in the {MFS}}.  \jt{Numerical
  Algorithms}  \bvol{72}~(1),  \pg{107--130}.

\bibitem[Cheng \& Hong(2020)]{cheng_overview_2020}
{\sc \au{Cheng, Alexander~H.D.} \& \au{Hong, Yongxing}} \yr{2020}  \at{An
  overview of the method of fundamental solutions—{Solvability}, uniqueness,
  convergence, and stability}.  \jt{Engineering Analysis with Boundary
  Elements}  \bvol{120},  \pg{118--152}.

\bibitem[Choi {\em et~al.\/}(2003)Choi, Westin \& Breuer]{choi_apparent_2003}
{\sc \au{Choi, Chang-Hwan}, \au{Westin, K. Johan~A.} \& \au{Breuer,
  Kenneth~S.}} \yr{2003}  \at{Apparent slip flows in hydrophilic and
  hydrophobic microchannels}.  \jt{Physics of Fluids}  \bvol{15}~(10),
  \pg{2897--2902}.

\bibitem[Epstein(1924)]{epstein_resistance_1924}
{\sc \au{Epstein, Paul~S.}} \yr{1924}  \at{On the {Resistance} {Experienced} by
  {Spheres} in their {Motion} through {Gases}}.  \jt{Physical Review}
  \bvol{23}~(6),  \pg{710--733}.

\bibitem[Falk {\em et~al.\/}(2010)Falk, Sedlmeier, Joly, Netz \&
  Bocquet]{falk_molecular_2010}
{\sc \au{Falk, Kerstin}, \au{Sedlmeier, Felix}, \au{Joly, Laurent}, \au{Netz,
  Roland~R.} \& \au{Bocquet, Lydéric}} \yr{2010}  \at{Molecular {Origin} of
  {Fast} {Water} {Transport} in {Carbon} {Nanotube} {Membranes}:
  {Superlubricity} versus {Curvature} {Dependent} {Friction}}.  \jt{Nano
  Letters}  \bvol{10}~(10),  \pg{4067--4073}.

\bibitem[Gad-el Hak(1999)]{gad-el-hak_fluid_1999}
{\sc \au{Gad-el Hak, Mohamed}} \yr{1999}  \at{The {Fluid} {Mechanics} of
  {Microdevices}—{The} {Freeman} {Scholar} {Lecture}}.  \jt{Journal of Fluids
  Engineering}  \bvol{121}~(1),  \pg{5--33}.

\bibitem[Happel \& Brenner(1983)]{happel_low_1983}
{\sc \au{Happel, J.} \& \au{Brenner, H.}} \yr{1983} {\em Low {Reynolds} number
  hydrodynamics: with special applications to particulate media\/}.
  \publ{Springer Netherlands}.

\bibitem[Holt {\em et~al.\/}(2006)Holt, Park, Wang, Stadermann, Artyukhin,
  Grigoropoulos, Noy \& Bakajin]{holt_fast_2006}
{\sc \au{Holt, Jason~K.}, \au{Park, Hyung~Gyu}, \au{Wang, Yinmin},
  \au{Stadermann, Michael}, \au{Artyukhin, Alexander~B.}, \au{Grigoropoulos,
  Costas~P.}, \au{Noy, Aleksandr} \& \au{Bakajin, Olgica}} \yr{2006}  \at{Fast
  {Mass} {Transport} {Through} {Sub}-2-{Nanometer} {Carbon} {Nanotubes}}.
  \jt{Science}  \bvol{312}~(5776),  \pg{1034--1037}.

\bibitem[Karageorghis(2009)]{karageorghis_practical_2009}
{\sc \au{Karageorghis, A.}} \yr{2009}  \at{A {Practical} {Algorithm} for
  {Determining} the {Optimal} {Pseudo}-{Boundary} in the {Method} of
  {Fundamental} {Solutions}}.  \jt{Advances in Applied Mathematics and
  Mechanics}  \bvol{1}~(4),  \pg{510--528}.

\bibitem[Karniadakis \& Beşkök(2002)]{karniadakis_micro_2002}
{\sc \au{Karniadakis, George} \& \au{Beşkök, Ali}} \yr{2002} {\em Micro
  flows: fundamentals and simulation\/}.  \publ{New York: Springer}.

\bibitem[Keh \& Chang(2008)]{keh_slow_2008}
{\sc \au{Keh, H.J.} \& \au{Chang, Y.C.}} \yr{2008}  \at{Slow motion of a slip
  spheroid along its axis of revolution}.  \jt{International Journal of
  Multiphase Flow}  \bvol{34}~(8),  \pg{713--722}.

\bibitem[Lauga {\em et~al.\/}(2007)Lauga, Brenner \&
  Stone]{tropea_microfluidics_2007}
{\sc \au{Lauga, Eric}, \au{Brenner, Michael} \& \au{Stone, Howard}} \yr{2007}
  \at{Microfluidics: {The} {No}-{Slip} {Boundary} {Condition}}.  \bt{In {\em
  Springer {Handbook} of {Experimental} {Fluid} {Mechanics}\/} (ed. \ed{Cameron
  Tropea, Alexander~L. Yarin \& John~F. Foss})},  \pg{pp. 1219--1240}.
  \publ{Berlin, Heidelberg: Springer Berlin Heidelberg}.

\bibitem[Lauga \& Stone(2003)]{lauga_effective_2003}
{\sc \au{Lauga, Eric} \& \au{Stone, Howard~A.}} \yr{2003}  \at{Effective slip
  in pressure-driven {Stokes} flow}.  \jt{Journal of Fluid Mechanics}
  \bvol{489},  \pg{55--77}.

\bibitem[Li {\em et~al.\/}(2006)Li, Chu \& Chen]{li_partially_2006}
{\sc \au{Li, W.-L.}, \au{Chu, H.-M.} \& \au{Chen, M.-D.}} \yr{2006}  \at{The
  partially wetted bearing—extended {Reynolds} equation}.  \jt{Tribology
  International}  \bvol{39}~(11),  \pg{1428--1435}.

\bibitem[Lighthill(1952)]{lighthill_squirming_1952}
{\sc \au{Lighthill, M.~J.}} \yr{1952}  \at{On the squirming motion of nearly
  spherical deformable bodies through liquids at very small reynolds numbers}.
  \jt{Communications on Pure and Applied Mathematics}  \bvol{5}~(2),
  \pg{109--118}.

\bibitem[Lockerby(2022)]{lockerby_integration_2022}
{\sc \au{Lockerby, D.~A.}} \yr{2022}  \at{Integration over discrete closed
  surfaces using the {Method} of {Fundamental} {Solutions}}.  \jt{Engineering
  Analysis with Boundary Elements}  \bvol{136},  \pg{232--237}.

\bibitem[Lockerby \& Collyer(2016)]{lockerby_fundamental_2016}
{\sc \au{Lockerby, D.~A.} \& \au{Collyer, B.}} \yr{2016}  \at{Fundamental
  solutions to moment equations for the simulation of microscale gas flows}.
  \jt{Journal of Fluid Mechanics}  \bvol{806},  \pg{413--436}.

\bibitem[Lockerby \& Reese(2008)]{lockerby_modelling_2008}
{\sc \au{Lockerby, D.~A.} \& \au{Reese, J.~M.}} \yr{2008}  \at{On the modelling
  of isothermal gas flows at the microscale}.  \jt{Journal of Fluid Mechanics}
  \bvol{604},  \pg{235--261}.

\bibitem[Lockerby {\em et~al.\/}(2004)Lockerby, Reese, Emerson \&
  Barber]{lockerby_velocity_2004}
{\sc \au{Lockerby, D.~A.}, \au{Reese, J.~M.}, \au{Emerson, D.~R.} \&
  \au{Barber, R.~W.}} \yr{2004}  \at{Velocity boundary condition at solid walls
  in rarefied gas calculations}.  \jt{Physical Review E}  \bvol{70}~(1),
  \pg{017303}.

\bibitem[Maxwell(1879)]{maxwell_stresses_1879}
{\sc \au{Maxwell, J.C.}} \yr{1879}  \at{On stresses in rarified gases arising
  from inequalities of temperature}.  \jt{Philosophical Transactions of the
  Royal Society of London}  \bvol{170},  \pg{231--256}.

\bibitem[Nicholls {\em et~al.\/}(2012)Nicholls, Borg, Lockerby \&
  Reese]{nicholls_water_2012}
{\sc \au{Nicholls, William~D.}, \au{Borg, Matthew~K.}, \au{Lockerby, Duncan~A.}
  \& \au{Reese, Jason~M.}} \yr{2012}  \at{Water transport through (7,7) carbon
  nanotubes of different lengths using molecular dynamics}.  \jt{Microfluidics
  and Nanofluidics}  \bvol{12}~(1-4),  \pg{257--264}.

\bibitem[Oberbeck(1876)]{oberbeck_ueber_1876}
{\sc \au{Oberbeck, A}} \yr{1876}  \at{Ueber stationäre
  {Flüssigkeitsbewegungen} mit {Berücksichtigung} der inneren {Reibung}.}
  \jt{Journal für die reine und angewandte Mathematik (Crelles Journal)}
  \bvol{1876}~(81),  \pg{62--80}.

\bibitem[Payne \& Pell(1960)]{payne_stokes_1960}
{\sc \au{Payne, L.~E.} \& \au{Pell, W.~H.}} \yr{1960}  \at{The {Stokes} flow
  problem for a class of axially symmetric bodies}.  \jt{Journal of Fluid
  Mechanics}  \bvol{7}~(4),  \pg{529--549}.

\bibitem[Persson \& Strang(2004)]{persson_simple_2004}
{\sc \au{Persson, Per-Olof} \& \au{Strang, Gilbert}} \yr{2004}  \at{A {Simple}
  {Mesh} {Generator} in {MATLAB}}.  \jt{SIAM Review}  \bvol{46}~(2),
  \pg{329--345}.

\bibitem[Qin {\em et~al.\/}(2011)Qin, Yuan, Zhao, Xie \&
  Liu]{qin_measurement_2011}
{\sc \au{Qin, Xingcai}, \au{Yuan, Quanzi}, \au{Zhao, Yapu}, \au{Xie, Shubao} \&
  \au{Liu, Zhongfan}} \yr{2011}  \at{Measurement of the {Rate} of {Water}
  {Translocation} through {Carbon} {Nanotubes}}.  \jt{Nano Letters}
  \bvol{11}~(5),  \pg{2173--2177}.

\bibitem[Rothstein(2010)]{rothstein_slip_2010}
{\sc \au{Rothstein, Jonathan~P.}} \yr{2010}  \at{Slip on {Superhydrophobic}
  {Surfaces}}.  \jt{Annual Review of Fluid Mechanics}  \bvol{42}~(1),
  \pg{89--109}.

\bibitem[Shahdhaar {\em et~al.\/}(2020)Shahdhaar, Yadawad, Khamari \&
  Behera]{shahdhaar_numerical_2020}
{\sc \au{Shahdhaar, M.~A.}, \au{Yadawad, S.~S.}, \au{Khamari, D.~S.} \&
  \au{Behera, S.~K.}} \yr{2020}  \at{Numerical investigation of slip flow
  phenomenon on performance characteristics of gas foil journal bearing}.
  \jt{SN Applied Sciences}  \bvol{2}~(10),  \pg{1677}.

\bibitem[Sherman(1990)]{sherman_viscous_1990}
{\sc \au{Sherman, Frederick~S.}} \yr{1990} {\em Viscous flow\/}.  \publ{New
  York St Louis Paris [etc.]: McGraw-Hill}.

\bibitem[Singh {\em et~al.\/}(1984)Singh, Rao \& Majumdar]{singh_effect_1984}
{\sc \au{Singh, K.~C.}, \au{Rao, N.~S.} \& \au{Majumdar, B.C.}} \yr{1984}
  \at{Effect of {Slip} {Flow} on the {Steady}-{State} {Performance} of
  {Aerostatic} {Porous} {Journal} {Bearings}}.  \jt{Journal of Tribology}
  \bvol{106}~(1),  \pg{156--162}.

\bibitem[Sone(2002)]{sone_kinetic_2002}
{\sc \au{Sone, Yoshio}} \yr{2002} {\em Kinetic theory and fluid dynamics\/}.
  \publ{Boston: Birkhäuser}, oCLC: ocm49726401.

\bibitem[Torrilhon(2016)]{torrilhon_modeling_2016}
{\sc \au{Torrilhon, Manuel}} \yr{2016}  \at{Modeling {Nonequilibrium} {Gas}
  {Flow} {Based} on {Moment} {Equations}}.  \jt{Annual Review of Fluid
  Mechanics}  \bvol{48}~(1),  \pg{429--458}.

\bibitem[Zhang {\em et~al.\/}(2011)Zhang, Zhou \& Meng]{zhang_performance_2011}
{\sc \au{Zhang, W.-M.}, \au{Zhou, J.-B.} \& \au{Meng, G.}} \yr{2011}
  \at{Performance and stability analysis of gas-lubricated journal bearings in
  {MEMS}}.  \jt{Tribology International}  \bvol{44}~(7-8),  \pg{887--897}.

\end{thebibliography}
\end{document}